\documentclass[10pt,journal,twoside]{IEEEtran}
\usepackage{multirow}
\usepackage{amssymb}
\usepackage[dvips]{graphicx}
\usepackage{amsmath}
\usepackage{amsthm}
\usepackage{array}
\usepackage{latexsym,bm}
\usepackage{color}
\usepackage{makecell}
\usepackage{subfigure}
\usepackage{longtable}
\usepackage{accents}
\usepackage{cite}
\usepackage{enumerate}
\usepackage{arydshln}

\usepackage[ruled]{algorithm}
\usepackage{algorithmic}
\usepackage{stfloats}
\usepackage{diagbox}
\makeatletter
\def\widebar{\accentset{{\cc@style\underline{\mskip10mu}}}}
\def\Widebar{\accentset{{\cc@style\underline{\mskip8mu}}}}
\makeatother

\allowdisplaybreaks
\theoremstyle{plain}

\theoremstyle{definition}
\theoremstyle{definition} 

\begin{document}

\title{On the Performance of LoRa Empowered Communication for Wireless Body Area Networks
\thanks{This research has been partially supported by the NSF of China grant number 62071129, and partially supported by the Academy of Finland, 6G Flagship program under Grant 346208.}
\thanks{M. Zhang and G. Cai are with the School of Information Engineering, Guangdong University of Technology, China (e-mail: 2112103025@mail2.gdut.edu.cn, caiguofa2006@gdut.edu.cn).}
\thanks{Z. Xu is with the School of Ocean Information Engineering, Jimei University, Xiamen, China (e-mail: xzpxmu@gmail.com).}
\thanks{J. He is with the Technology Innovation Institute, 9639 Masdar City, Abu Dhabi, United Arab Emirates, and he is also  with the Centre for Wireless Communications, University of Oulu, 90014 Oulu, Finland (E-mail: jiguang.he@tii.ae).}
\thanks{M. Juntti is with Centre for Wireless Communications, University of Oulu, P.O.Box 4500, FI-90014, Finland (email: markku.juntti@oulu.fi).}
}

\author{Minling~Zhang,~Guofa~Cai,~Zhiping~Xu,~Jiguang~He,~Markku~Juntti,~\IEEEmembership{Fellow,~IEEE}}

\maketitle
\begin{abstract}

To remotely monitor the physiological status of the human body, long range (LoRa) communication has been considered as an eminently suitable candidate for wireless body area networks (WBANs).
Typically, a Rayleigh-lognormal fading channel is encountered by the LoRa links of the WBAN.
In this context, we characterize the performance of the LoRa system in WBAN
scenarios with an emphasis on the physical (PHY) layer and
medium access control (MAC) layer in the face of Rayleigh-lognormal
fading channels and the same spreading factor interference.
Specifically, closed-form approximate bit error probability (BEP) expressions are derived for the LoRa system.
The results show that increasing the SF and reducing the interference efficiently mitigate the shadowing effects.
Moreover, in the quest for the most suitable MAC protocol for LoRa based WBANs, three MAC protocols are critically appraised, namely the pure ALOHA, slotted ALOHA, and carrier-sense multiple access.
The coverage probability, energy efficiency, throughput, and system delay of the three MAC protocols are analyzed in Rayleigh-lognormal fading channel.
Furthermore, the performance of the equal-interval-based and equal-area-based schemes is analyzed to guide the choice of the SF.
Our simulation results confirm the accuracy of the mathematical analysis and provide some useful insights for the future design of LoRa based WBANs.

\end{abstract}


\section{Introduction} \label{sect:review}
Healthcare Internet of Thing (HIoT) is a highly efficient and convenient way to provide intelligent
diagnosis, treat and disease management, and anticipate risks to patient health \cite{2020The}. Especially, with the rapid spread of COVID-19, HIoT has been adopted to monitor the significant physiological information in the human body \cite{2020A}. Hence, it is important for HIoT to ensure the low-power, highly reliable, and long-range information transmission.

As an important technology of HIoT, wireless body area network (WBAN) consists of multiple interconnected low-power and resource-constrained sensor devices (e.g., worn, implanted, embedded, swallowed, etc.) that are located in-on-and-around the human body, where these sensor devices are adopted for sensing and data communication \cite{2012IEEE,2020T,2012Body,9857866}.
According to \cite{9585057}, in a HIoT system, physiological signals from each sensor device are transmitted to a hub via a WBAN, and then the collected data is forwarded to the service center.
Thus, these information can be provided to the hospitals or clinics. However, due to the limited transmission distance of the WBAN, it is not suitable for long-distance transmission.

Long range (LoRa) modulation, as the physical (PHY) layer of LoRa network, is a chirp spread spectrum (CSS) based modulation,  which can  achieve low-power and long-range transmission \cite{9709288}.
In the LoRa system, the receiver (usually a gateway) can decode the signal a few
decibels below the noise floor.
The range and data rate of LoRa communication can be adjusted by different spreading factors (SFs).
In addition, LoRa gateway can correctly receive two overlapping signals over the same channel, as long as their SFs are different.
Due to these advantages, LoRa modulation was applied in WBAN applications for the off-body communications \cite{2021Performance}.
In \cite{2019Monitoring} and \cite{2018WE}, LoRa sensors were worn on the body to measure the vital signs and environmental data.
In \cite{9709339}, a low-power healthcare WBAN platform based on LoRa (HeaLoRa) was proposed for monitoring physiological parameters.
In \cite{9135635}, LoRa sensors were placed inside the animal body to  monitor the important parameters and communicates with a distant gateway.
In such a scenario, the in-to-out-body path loss (PL) was characterized for the first time at $868$ MHz.
In \cite{9142371}, a low-power LoRa link was built between a fixed base station and a mobile user which is worn on the body.
The test results showed that wide ranges can be easily and reliably achieved for off-body LoRa communication links.
In \cite{9169667}, the LoRa system was introduced for search and rescue applications in mountain areas.
Through the above discussions, it can be found that  theoretical analyses of the LoRa system for in-body and off-body communications are still lacking, which limits its further design.
In \cite{8392707}, the bit error probability (BEP) expression of the LoRa system was derived over a Rayleigh fading channel.
In \cite{7803607}, success probability analysis of the LoRa system was performed over a Rayleigh fading channel.
However, for LoRa comminution links in the WBAN, a Rayleigh-lognormal fading channel should be applied according to \cite{9135635,9142371,9169667}.

LoRa wide area network (LoRaWAN) is a medium access control (MAC) protocol designed to run LoRa modulation \cite{9773149}.
According to the LoRaWAN characteristics, the devices generally adopt a pure ALOHA (P-ALOHA) mechanism to access channels \cite{9241021}, where the packets access the channel randomly.
In \cite{7803607}, scalability analysis of the LoRa network for P-ALOHA mechanism
was performed over Rayleigh fading channels.
Although P-ALOHA has simple implementation, it is easy to cause packet collisions.
As a consequence, it brings the same SF (defined as \textit{co-SF}) interference, thus significantly  affecting the scalability of the LoRa network.
In \cite{8430542}, the equal-interval-based (EIB) and equal-area-based (EAB) schemes were used to compare and analyze the scalability of LoRa networks.
The results show that the scalability of LoRa network is significantly affected by the SF allocation schemes.
To enhance the scalability of LoRa network, the slotted ALOHA (S-ALOHA) and carrier-sense multiple access (CSMA) mechanisms were extensively studied in \cite{2019Slotted,8422800,8361706}.
In particular, P-ALOHA, S-ALOHA, and non-persistent CSMA (NP-CSMA) protocols for LoRa network was analyzed over Rayleigh fading channels \cite{9018210}.

Most of the aforementioned studies on PHY and MAC layers of LoRa networks focus on Rayleigh fading channels, and the impact of large-scale fading is ignored.
In the LoRa based WBAN, large-scale fading should be taken into account.
Motivated by the preliminary investigation in \cite{8392707,7803607}, and \cite{9018210},
we propose a more comprehensive and tractable framework of performance analysis on LoRa system for WBAN subject to the Rayleigh-lognormal channels and the \textit{co-SF} interference from the perspective of the PHY and MAC layers. The main contributions of this paper are summarized as follows:


\begin{enumerate}
	\item
	A LoRa based WBAN is put forward, which consists of an enormous number of LoRa end-devices (EDs) from the WBANs and a gateway.
	In such a scenario, the Rayleigh-lognormal channel and the \textit{co-SF} interference are jointly considered.

	\item
	The closed-form BEP expression of the LoRa system for PHY transmission is derived under the Rayleigh-lognormal channel and \textit{co-SF} interference.
	Furthermore, we investigate the impact of SF, the shadowing standard deviation  and the signal-to-interference ratio (SIR) on the BEP.
	It is shown that although the BEP of the proposed system is significantly affected by the shadowing standard deviation, the system is capable of mitigating the shadowing effects by increasing the SF and SIR.

	\item
	To find the most appropriate MAC protocol for the LoRa based WBAN,  the P-ALOHA, S-ALOHA, and NP-CSMA protocols are comprehensively investigated and compared.
   In addition, to elaborate on the choice of the SFs, the EIB and EAB schemes are considered.
  Moreover, we analyse the coverage probability, energy efficiency, throughput, and system delay of LoRa networks relying on the above-mentioned protocols and SF allocation schemes for the transmission.
Furthermore, we also analyze the impact of some important parameters, i.e., network radius  and the average number of EDs, on the system performance.
	
\end{enumerate}

The remainder of this paper is organized as follows.
Section~\ref{sect:system_model} introduces the system model of LoRa based WBAN from the perspective of PHY and MAC layers.
In Section~\ref{sect:Performance_Analysis}, the performance of the LoRa based WBAN is analyzed in terms of BEP and some metrics of scalability.
Section~\ref{sect:Results_and_Discussion} presents the results and discussions.
Finally, Section~\ref{sect:Conclusion} concludes this paper.

\section{System Model} \label{sect:system_model}

\begin{figure}[t]
	\center
	\includegraphics[width=3.6in,height=1.8in]{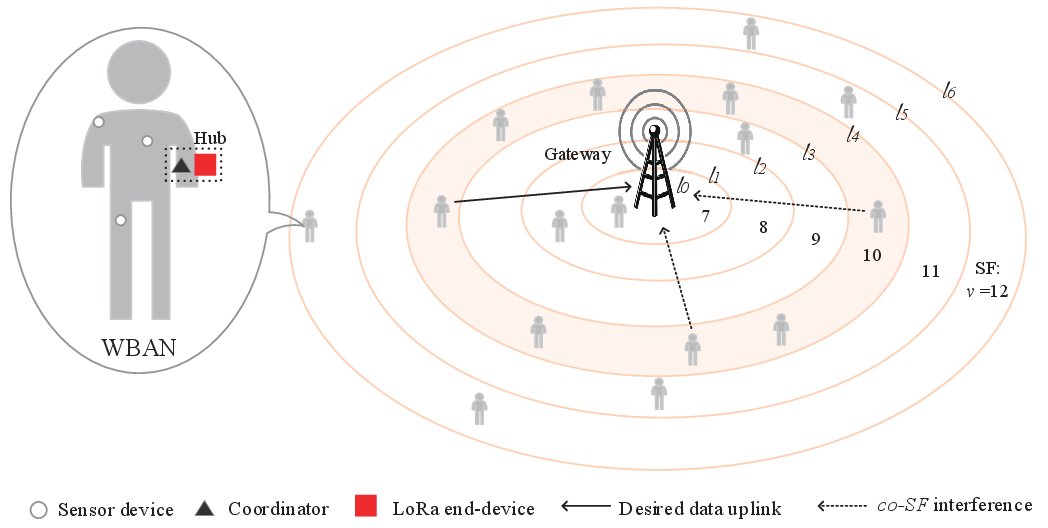}
	\caption{ Uplink transmission for the LoRa based WBAN, where the SF is allocated based on annuli.}
	\label{fig:fig1}
	\vspace{-0.45cm}
\end{figure}

Uplink transmission for the LoRa based WBAN is shown in Fig.~\ref{fig:fig1}, which includes a large number of LoRa EDs from the WBANs and a gateway.
For each WBAN, there are many sensor devices and a hub, where the hub has a coordinator and a LoRa ED.
The coordinator collects the physiological data from the sensor devices.
The LoRa EDs send the collected physiological data to the gateway.
In Fig.~\ref{fig:fig1}, we consider the interference caused by the signals generated by the activated EDs with the same SF. It should be noted that the signals for different SFs are perfectly orthogonal.
The EDs are uniformly distributed within a circle of radius $R$ (km), which can be described by \textit{homogeneous Poisson point process} (PPP) $\Phi$ with intensity $\lambda$ ($\lambda$\textgreater0).
We assume that $\mathcal{V} \subseteq \mathbb{R}^2$ (2-dimensional Euclidean space) is the disk of radius $R$, of which the area is $V = \left| \mathcal{V} \right|=\pi {R^2}$. The total number of EDs within the disk is $\mathcal{N}$, which is a random variable following a Poisson distribution with mean $\overline{\mathcal{N}}=\lambda V $. The probability density function (PDF) of the distance between the randomly selected ED and the gateway can be expressed as ${f_{\rm ED}}\left( x \right) = \frac{{2\pi x}}{{\pi {R^2}}}$, $0 \le x \le R$. The disk is divided into $K$ parts, each of which corresponds to a different SF $\nu$, where $\nu \in \Psi=\{7,8,9,10,11,12\}$ and $ K = \left| \Psi  \right| $ represents the number of annuli with $|\cdot|$ being the cardinality of a set.
The inner and outer diameters of the $j$-th ($1 \le j \le  K$) annulus are defined as ${l_{j-1}}$ and ${l_j}$, respectively.

\subsection{Signal Model for  LoRa Physical Layer} \label{sect:Signal Model for  LoRa Physical Layer}
LoRa modulation is based on CSS. If the bandwidth of chirp signal is $B$, a LoRa sample is sent every elapsed time $T = \frac{1}{B}$. Besides, LoRa modulation is realized by spreading the frequency change of the chirp signal over ${2^{\nu}}$ samples within a symbol duration ${T_s} = {2^{\nu}} T$.
Each symbol $s_q$ can carry $\nu$ bits of information and $s_q = q$, $q \in \left\{ {0,1,2, \cdot  \cdot  \cdot ,{2^{\nu}} - 1} \right\}$.
The chirp signal ${\varpi _q}\left( {mT} \right) $ can be expressed as
\begin{align}
\label{eq:chirp_signal}
{\varpi _q}\left( {mT} \right) = \sqrt {\frac{{{1}}}{{{2^{\nu}}}}}  {e^{j2\pi \left[ {\left( {q + m} \right)\bmod {2^{\nu}}} \right]\frac{m}{{{2^{\nu}}}}}},
\end{align}
where $m= 0$, $1$, $2$, $\cdots$, ${{2^{\nu}} - 1}$ indicates the symbol index at time $mT$.
Moreover, the transmitted discrete-time LoRa baseband signal can be expressed as\cite{8392707}
\begin{align}
\label{eq:transmit_signal}
{\omega _q}\left( {mT} \right) &= \sqrt {{E_s}}  {\varpi _q}\left( {mT} \right) \nonumber\\
&= \sqrt {\frac{{{E_s}}}{{{2^{\nu}}}}} {e^{j2\pi \left[ {\left( {q + m} \right)\bmod {2^{\nu}}} \right]\frac{m}{{{2^{\nu}}}}}},
\end{align}
where $E_s$ is the energy per symbol.

From \cite{8392707}, based on the orthogonality of chirp signals with different offsets, the cross-correlation property of the LoRa demodulator is written as
\begin{align}
\label{eq:cross-correlation}
{C_{\iota ,q}} = \sum\limits_{m = 0}^{{2^\nu } - 1} {{\varpi _q }} \left( {mT} \right) \varpi _{_\iota}^*\left( {mT} \right) = \left\{ {\begin{array}{*{20}{c}}
	1,&{\iota = q }\\
	0,&{\iota \ne q }
	\end{array}} \right.,
\end{align}
where the chirp signal ${\varpi _\iota}\left( {mT} \right) $ is used to transmit a symbol $s_\iota = \iota$ $, \iota \in \left\{ {0,1,2,\cdots,{2^\nu } - 1} \right\}$, and $\varpi _{_\iota}^*\left( {mT} \right)$ is the complex conjugate of the chirp signal ${\varpi _\iota}\left( {mT} \right)$.


Due to the human body shadows and environmental hindrances, the communication channel can be modeled as a Rayleigh-lognormal fading channel \cite{2014Performance}.
Since the LoRa transmitters generally work asynchronously, collisions always occur between different signals.
At the gateway, the received signal can be represented as
\begin{align}
	\label{eq:receive_signal}
{r_q}\left( {mT} \right)\! = \!{h_1}  {\omega _q}\left( {mT} \right)\! + \!\sum\limits_{k = 2}^\mathcal{N}  {\chi _k^\nu  {h_k}  {\omega _{I,k}}\left( {mT} \right)} \! +\! {\phi _q}\left( {mT} \right),
\end{align}
where $h_1$ and $h_k$ represent the channel coefficient between the desired ED and the gateway, and that between the interfering ED and the gateway, respectively.
$\chi _k^{\nu}$ is the output of the indicator function, where $\chi _k^{\nu} = 1$ if the $k$-th interfering node within the same SF annulus as the desired ED, otherwise $\chi _k^{\nu} = 0$.
${\omega_q}\left( {mT} \right)$ is the desired signal, ${\omega _{I,k}}\left( {mT} \right)$ is the interfering signal from the $k$-th interfering node, and ${\phi_q}\left( {mT} \right)$ is complex additive white Gaussian noise (AWGN) with zero-mean and variance ${N_0}$.

Due to the capture effect of LoRa \cite{8886735}, we only need to consider the strongest interfering signal in this paper.
Let ${h_{{\hat{k}}}}$ denote the channel coefficient between the strongest interfering ED and the gateway, and ${\omega  _{I,\hat{k}}}\left( {mT} \right)$ denotes the strongest interfering signal, (\ref{eq:receive_signal}) can be expressed as
\begin{align}
\label{eq:receive_signal2}
{r_q}\left( {mT} \right) = {h_1}  {\omega _q}\left( {mT} \right) + {h_{{\hat{k}}}}  {\omega _{{I,\hat{k}}}}\left( {mT} \right) + {\phi _q}\left( {mT} \right).
\end{align}


The log-distance PL model between an ED and a gateway is given by
\begin{equation}
\begin{aligned}
\label{eq:pathloss}
{P_L}(d) = {P_L}({d_0}) + 10 n {\log _{10}}\frac{d}{{{d_0}}} + S[dB],
\end{aligned}
\end{equation}
where $P_L(d_0)$ is the PL in dB at the reference distance $d_0 = 1$ m, $d$ is the distance between the ED and the gateway, $n$ is the PL exponent and  $S \! \sim \!  \mathcal{N}\! (0,{\sigma _{dB}^2})$ is a normally distributed random variable that represents the shadowing effect.

Without loss of generality, combining the fading with the shadowing, the instantaneous normalized channel power ${\left| h \right|^2}$ can be given by
\begin{align}
\label{eq:channel_gain}
{\left| h \right|^2} &= {\left| {{h_\mathrm{ray}}} \right|^2}  {\left| {{h_\mathrm{log }}} \right|^2} \nonumber\\
&= {\left| {{h_\mathrm{ray}}} \right|^2}  {10^{ - \frac{{{P_L}(d)[dB]}}{{10}}}} \nonumber\\
&= {\left| {{h_\mathrm{ray}}} \right|^2}  {10^{ - \frac{{{P_L}({d_0}) + 10n{{\log }_{10}}\frac{d}{{{d_0}}}}}{{10}}}} \times {10^{ - \frac{S}{{10}}}} \nonumber\\
&= {\left| {{h_\mathrm{ray}}} \right|^2}H g(d) \nonumber\\
&= {\left| {{\hbar }} \right|^2} g(d),
\end{align}
where ${\left| {{h_\mathrm{ray}}} \right|^2}$ is the channel power gain of a Rayleigh channel modelling as an exponential random variable with mean one, i.e., ${\left| {{h_\mathrm{ray}}} \right|^2}$ $\sim$ exp (1).
${\left| {{h_\mathrm{log }}} \right|^2}$ represents the channel power gain of a log-normal channel.
$H = {10^{ - \frac{S}{{10}}}}$ follows a log-normal distribution with mean ${\mu _H} = 0$ and standard deviation ${\sigma _H} = \frac{{\ln 10}}{{10}}{\sigma _{dB}}$.
${g\left( d \right)}= {10^{ - \frac{{{P_L}\left( {{d_0}} \right) + 10n{{\log }_{10}}\frac{d}{{{d_0}}}}}{{10}}}}$ is the path loss attenuation, and ${\left| \hbar \right|^2} = {\left| {{h_\mathrm{ray}}} \right|^2}H$ follows a Rayleigh-lognormal distribution.
Hence, (\ref{eq:receive_signal2}) can be further written as
\begin{align}
\label{eq:receive_signal3}
{r_q}\left( {mT} \right) = &\sqrt {{{\left| {\hbar_1} \right|}^2}g\left( {{d_1}} \right)}  {\omega _q}\left( {mT} \right) \nonumber\\
&+ \sqrt {{{\left| {\hbar_{{\hat{k}}}} \right|}^2}g\left( {{d_{{\hat{k}}}}} \right)}  {\omega _{{I,\hat{k}}}}\left( {mT} \right) + {\phi _q}\left( {mT} \right),
\end{align}
where ${\hbar_1}$ and ${\hbar_{\hat{k}}}$ represent the channel gain of Rayleigh-lognormal channel between the desired ED and the gateway, and that between the strongest interfering ED and the gateway, respectively.
$d_1$ and $d_{\hat{k}}$ denote the distance between the desired ED and the gateway, and that between the strongest interfering ED and the gateway, respectively.

Besides, the correlator output of the LoRa demodulator is given by
\begin{align}
\label{eq:correlator_output}
{{\rm O}_\iota }& = \sum\limits_{m = 0}^{{2^\nu } - 1} {{r_q}\left( {mT} \right) \varpi _{_\iota }^*} \left( {mT} \right) \nonumber\\
&= \left\{ {\begin{array}{*{20}{c}}
	{\sqrt {{\beta _1}{E_s}g({d_1})}  + \sqrt {{\beta _{\hat{k}}}{E_s}g({d_{\hat{k}}})}{{\Delta _\iota }}  + {\phi _\iota }\left( {mT} \right)},&{\iota  = q}\\
	{\sqrt {{\beta _{\hat{k}}}{E_s}g({d_{\hat{k}}})}{{\Delta _\iota }}  + {\phi _\iota }\left( {mT} \right)},&{\iota  \ne q}
	\end{array}} \right.,
\end{align}
where $\beta_1 = {{\left| \hbar_1 \right|}^2}$, $\beta_{\hat{k}} = {{\left| {\hbar_{\hat{k}}} \right|}^2}$, $\phi_\iota \left( {mT} \right)$ is also complex AWGN with zero-mean and variance ${N_0}$, and ${{\Delta _\iota }}$ is the cross-correlation interference.

%

\begin{figure}[t]
	\center
	\includegraphics[width=3.5in,height=1.4in]{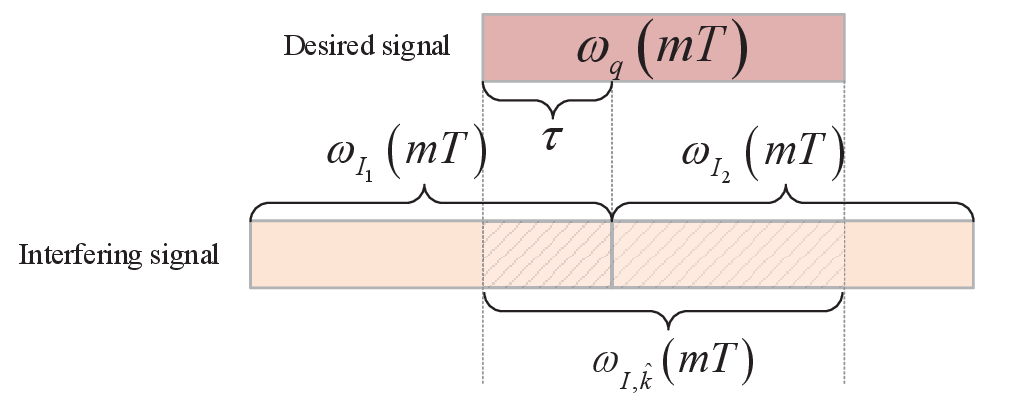}
	\caption{ Illustration of the relationship between the desired signal and the interfering signal.}
	\label{fig:fig2}
	\vspace{-0.45cm}
\end{figure}

Fig.~\ref{fig:fig2} shows the relationship between the desired signal and the interfering signal. According to \cite{8581011} and this figure, the interfering signal is defined as

\begin{align}
\label{eq:twopatrs_inter_signal}
{\omega _{{I,\hat{k}}}}\left( {mT} \right) = \left\{ {\begin{array}{*{20}{c}}
	{\frac{1}{{\sqrt \rho  }}{\omega _{{I_1}}}\left( {mT} \right)},&{0 \le m < \tau }\\
	{\frac{1}{{\sqrt \rho  }}{\omega _{{I_2}}}\left( {mT} \right)},&{\tau  \le m < {2^\nu }}
	\end{array}} \right.,
\end{align}
where ${\omega _{{I,\hat{k}}}}\left( {mT} \right)$ consists of two signals ${\omega _{{I_1}}}\left( {mT} \right)$ and ${\omega _{{I_2}}}\left( {mT} \right)$, $I_1, I_2 \in \left\{ {0,1,2, \cdot  \cdot  \cdot ,{2^{\nu}} - 1} \right\}$, $\rho$ represents the SIR, and it is assumed that a random offset $\tau$ occurs between the desired and interfering signal ${\omega _{{I_2}}}\left( {mT} \right)$.
Without loss of generality, $\tau$ is assumed to be uniformly distributed in $\left[ {0,{2^{\nu  - 1}}} \right]$, thus, it can ensure that the number of interfering samples from ${\omega _{{I_2}}}\left( {mT} \right)$ is larger than that from ${\omega _{{I_1}}}\left( {mT} \right)$.

Accordingly, one can assume $I_2 = 0$ to ensure that the last $2^\nu-\tau$ samples of ${\omega _q}\left( {mT} \right)$ are interfered by the signal ${\omega _0}\left( {mT} \right) = \sqrt {\frac{{{E_s}}}{{{2^\nu }}}}  {e^{j2\pi \left( m \right)\frac{m}{{{2^\nu }}}}}$ according to (\ref{eq:transmit_signal}).
Hence, ${{\Delta _\iota }}$ can be expressed as
\begin{align}
\label{eq:cross-c-inter}
{\Delta _\iota } = \frac{1}{{\sqrt \rho  }} \left( {\left. {{C_{\iota ,{I_1}}}} \right|_0^{\tau  - 1} + \left. {{C_{\iota ,0}}} \right|_\tau ^{{2^\nu } - 1}} \right),
\end{align}
where
\begin{align}
\label{eq:cc-inter_part1}
\left. {{C_{\iota ,{I_1}}}} \right|_0^{\tau  - 1} = \frac{1}{{{2^\nu }}} \sum\limits_{m = 0}^{\tau  - 1} {{e^{j2\pi \left( {\iota  - {I_1}} \right)\frac{m}{{{2^\nu }}}}}} ,
\end{align}
\begin{align}
\label{eq:cc-inter_part2}
\left. {{C_{\iota ,0}}} \right|_\tau ^{{2^\nu } - 1} = \frac{1}{{{2^\nu }}} \sum\limits_{m = \tau }^{{2^\nu } - 1} {{e^{j2\pi \iota \frac{m}{{{2^\nu }}}}}} .
\end{align}

Using the summation formula of the geometric sequence, the Euler's formula and the trigonometric formula, the magnitude of $\left. {{C_{\iota ,{I_1}}}} \right|_0^{\tau  - 1}$ and $\left. {{C_{\iota ,0}}} \right|_\tau ^{{2^\nu } - 1}$ can be expanded as
\begin{align}
\label{eq:cc-inter_part1_2}
\left| {\left. {{C_{\iota ,{I_1}}}} \right|_0^{\tau  - 1}} \right| = \frac{1}{{{2^\nu }}}  \left| {\frac{{\sin \left( {\pi \frac{{\left( {\iota  - {I_1}} \right)}}{{{2^\nu }}}\tau } \right)}}{{\sin \left( {\pi \frac{{\left( {\iota  - {I_1}} \right)}}{{{2^\nu }}}} \right)}}} \right|,
\end{align}
\begin{align}
\label{eq:cc-inter_part2_2}
\left| {\left. {{C_{\iota ,0}}} \right|_\tau ^{{2^\nu } - 1}} \right| = \frac{1}{{{2^\nu }}}  \left| {\frac{{\sin \left( {\pi \frac{\iota }{{{2^\nu }}}\left( {{2^\nu } - \tau } \right)} \right)}}{{\sin \left( {\pi \frac{\iota }{{{2^\nu }}}} \right)}}} \right|,
\end{align}
where $\left| {\left. {{C_{\iota ,{I_1}}}} \right|_0^{\tau  - 1}} \right|$ reaches its maximum value of $\frac{\tau }{{{2^\nu }}}$ when $\iota  = {I_1}$, and $\left| {\left. {{C_{\iota ,0}}} \right|_\tau ^{{2^\nu } - 1}} \right|$ reaches its maximum value of $\frac{{{2^\nu } - \tau }}{{{2^\nu }}}$ when $\iota  = 0$.

Let ${\mathcal{U} _\iota }$ be the upper bound of the magnitude of ${\Delta _\iota }$. It can be given by
\begin{align}
\label{eq:uppercc}
\left| {{\Delta _\iota }} \right|\!\! \le\! {\mathcal{U} _\iota }\! =\! \frac{1}{{{2^\nu }\sqrt \rho  }}  \left( {\left|\! {\frac{{\sin \left( {\pi \frac{{\left( {\iota  - {I_1}} \right)}}{{{2^\nu }}}\tau } \right)}}{{\sin \left( {\pi \frac{{\left( {\iota  - {I_1}} \right)}}{{{2^\nu }}}} \right)}}} \!\right|\! +\! \left|\! {\frac{{\sin \left( {\pi \frac{\iota }{{{2^\nu }}}\left( {{2^\nu } - \tau } \right)} \right)}}{{\sin \left( {\pi \frac{\iota }{{{2^\nu }}}} \right)}}}\! \right|} \right).
\end{align}

For each realization of $I_1$ and $\tau$, one can assume that ${\mathcal{U} _\iota }$ achieves the peak cross-correlation interference at some bin ${\tilde{\iota}} = \mathop{\arg\max}\limits_{\iota} \left( {\mathcal{U} _\iota } \right)$.
The cross-correlation interference terms are negligible when $\iota \ne {\tilde{\iota}}$.
Hence, ${\mathcal{U} _\iota }$ can be approximated as
\begin{align}
\label{eq:upperccappro}
{\mathcal{U} _\iota } \approx \left\{ {\begin{array}{*{20}{c}}
	{{\mathcal{U} _{\tilde{\iota}} }},&{\iota  = {\tilde{\iota}} }\\
	0,&{\iota  \ne {\tilde{\iota}}}
	\end{array}} \right..
\end{align}

Accordingly, since $\tau  \in \left[ {0,{2^{\nu  - 1}}} \right]$, the peak cross-correlation interference is most likely to occur at bin ${\tilde{\iota}} = 0$, and one has
\begin{align}
\label{eq:upperccappro_0}
{\mathcal{U} _{\tilde{\iota}} } \approx {\mathcal{U} _0} = \frac{1}{{{2^\nu }\sqrt \rho  }}  \left( {\left| {\frac{{\sin \left( {\pi \frac{{\left( {\iota  - {I_1}} \right)}}{{{2^\nu }}}\tau } \right)}}{{\sin \left( {\pi \frac{{\left( {\iota  - {I_1}} \right)}}{{{2^\nu }}}} \right)}}} \right| + \left( {{2^\nu } - \tau } \right)} \right).
\end{align}

Finally, combining (\ref{eq:correlator_output}), (\ref{eq:upperccappro}) and (\ref{eq:upperccappro_0}), the magnitude of the correlator output of the LoRa demodulator can be obtained as
\begin{align}
\label{eq:magnitude_output1}
&\Bigg| {\left. {{{\rm O}_\iota }} \right|q = 0} \Bigg| \nonumber\\
&\approx \left\{ {\begin{array}{*{20}{c}}
	{\left| {\sqrt {{\beta _1}{E_s}g\left( {{d_1}} \right)}  + \sqrt {{\beta _{\hat{k}}}{E_s}g\left( {{d_{\hat{k}}}} \right)} {\mathcal{U} _0} + {\phi _0}\left( {mT} \right)} \right|},&{\iota  = 0}\\
	{\left| {{\phi _\iota }\left( {mT} \right)} \right|},&{\iota  \ne 0}
	\end{array}} \right.,
\end{align}
\begin{align}
\label{eq:magnitude_output2}
\Bigg| {\left. {{{\rm O}_\iota }} \right|q \ne 0} \Bigg| \approx \left\{ {\begin{array}{*{20}{c}}
	{\left| {\sqrt {{\beta _{\hat{k}}}{E_s}g\left( {{d_{\hat{k}}}} \right)} {\mathcal{U} _0} + {\phi _0}\left( {mT} \right)} \right|},&{\iota  = 0}\\
	{\left| {\sqrt {{\beta _1}{E_s}g\left( {{d_1}} \right)}  + {\phi _q}\left( {mT} \right)} \right|},&{\iota  = q}\\
	{\left| {{\phi _\iota }\left( {mT} \right)} \right|},&{\iota  \ne 0,q}
	\end{array}} \right..
\end{align}

We select the index of the correlator output which attains the largest amplitude. Therefore, the detected symbol ${\tilde{s}_q}$ is written as
\begin{equation}
\begin{aligned}
\label{eq:detect}
{\tilde{s}_q} = \mathop{\arg\max}\limits_{\iota} \left( {\left| {{{\rm O}_\iota }} \right| } \right).
\end{aligned}
\end{equation}

\subsection{LoRa MAC Layer} \label{LoRa MAC Layer}

In the distance-based SF allocation scheme, a value of the SF is assigned for an ED based on the distance between the ED and the gateway.
Two types of distance-based SF allocation schemes, i.e., EIB and EAB, are considered in this paper.
The EIB scheme has equal-width of each annulus while the EAB scheme has equal-area of each annulus.
In both schemes, the entire network is divided into $K$ annuli. The parameters of the two schemes are shown in Table~\ref{table:allocation}.

\begin{table}[]	
	\begin{center}
		\caption{The Parameters for SF Allocation Schemes.}
		\begin{tabular}{ c c c }
			\hline
			Parameters  &  EIB & EAB \\
			\hline
			Width &  ${r_j} = \frac{R}{K} $  &  ${r_j} = R\sqrt {\frac{j}{K}} - R\sqrt {\frac{j-1}{K}}$  \\
			\hline
			Area &  $\left|{\mathcal{V} _j} \right| = \pi \left ( \frac{R}{K} \right )  ^{2} \left ( 2j-1 \right ) $  &  $ \left|{\mathcal{V}_j} \right|= \frac{{\pi {R^2}}}{K}$  \\		
			\hline
		\end{tabular}
		\label{table:allocation}	
	\end{center}
\end{table}

The traditional channel access mechanism for LoRa is P-ALOHA, which is easy to implement. However, due to its poor scalability, alternative access mechanisms can be considered.
The purpose of this paper is to investigate the performance of different access mechanisms, i.e., P-ALOHA, S-ALOHA and NP-CSMA.
According to \cite{9018210}, we can define the intensity $\lambda_u$ of the PPP of interferers $\Phi_u$ for $u\in \left\{ {\rm P\raisebox{0mm}{-}ALOHA},{\rm S\raisebox{0mm}{-}ALOHA},{\rm NP\raisebox{0mm}{-}CSMA} \right\}$.

\subsubsection{P-ALOHA} \label{sect:P-ALOHA}
For P-ALOHA, one can express the intensity of the interferers as
\begin{align}
\label{eq:lamdaPALOHA}
{\lambda _ {\rm {P\raisebox{0mm}{-}ALOHA}} } = 2\alpha \lambda,
\end{align}
where '2' means the vulnerability time of P-ALOHA is twice of the message time-on-air (ToA) and $\alpha$ is the duty cycle constraint.

\subsubsection{S-ALOHA} \label{sect:S-ALOHA}
In the S-ALOHA protocol, the collision is divided into intra-slot collision and inter-slot collision, where the inter-slot collision is divided into the collision with the previous time slot and that with the next time slot.
Hence, one can obtain the intensity of the interferers as

\begin{align}
\label{eq:lamdaSALOHA}
{\lambda _ {\rm {S\raisebox{0mm}{-}ALOHA}} } = \left( {1 + \frac{{{T_g}}}{{T_o }}} \right){p_s}  \alpha \lambda ,
\end{align}
where $T_g$ is the guard interval, ${T_o }$ is the value of ${\rm ToA}$,
${p_s} = 1 + Q\left( {\frac{{{T_g} + {T_p} - 5{T_s}}}{{\sqrt 2 {\sigma _{te}}}}} \right)+ Q\left( {\frac{{{T_g}}}{{\sqrt 2 {\sigma _{te}}}}} \right)$ is the total probability of collisions,
$Q\left( x \right) = \frac{1}{{\sqrt {2\pi } }}\int_x^\infty  {{e^{ - \frac{{{\varsigma ^2}}}{2}}}} d\varsigma $ is the $Q$-function,
$T_p$ is the message preamble duration, and $\sigma_{te}$ is the standard deviation of packet lengths.

\begin{figure*}
	\begin{align}
	\label{eq:timeonair}
	{T_o } = {T_s} \times \left[ {{N_{sp}} + 4.25 + 8 + \Bigg \lceil   \left( {\frac{{\max \left( {8 {N_{bp}} + {N_{CRC}} - 4  \nu + 8 + {N_{sh}},0} \right)}}{{4  \left( {\nu - 2} \right)}}} \right) \Bigg \rceil \times \left( {cr + 4} \right)} \right],
	\end{align}
	{\noindent} \rule[-10pt]{18cm}{0.05em}
\end{figure*}

Since we consider low data rate optimization mode, ${T_o }$ is given in (\ref{eq:timeonair}), where ${N_{bp}}$ is the payload size for the message, ${N_{sp}}$ is the length of the message preamble,
${N_{CRC}} = 16$ if cyclic redundancy check (CRC) is activated, otherwise $N_{CRC} = 0$.
${N_{sh}} = 20$ is for explicit header, ${N_{sh}} = 0$ is for implicit header, and $cr = 1$, $2$, $3$ or $4$ is associated with coding rate $CR= 4/5$, $4/6$, $4/7$ or $4/8$.


\subsubsection{NP-CSMA} \label{sect:NP-CSMA}
According to \cite{9018210}, one has
\begin{align}
\label{eq:lamdaCSMA}
{\lambda _{\rm NP\raisebox{0mm}{-}CSMA}} = \left( {2 - \frac{{{T_p} - 5{T_s}}}{{T_o }}} \right)\left( {1 - \Xi } \right)\frac{{1 - {e^{ - E\left( {n_A } \right)}}}}{{E\left( {n_A } \right)}}p\lambda,
\end{align}
where ${\frac{{{T_p} - 5{T_s}}}{{T_o }}}$ is the reduction of vulnerability time due to the nature of LoRa transmission, $p$ is the probability for an ED to be granted access to the channel ($p>\alpha$),
${E\left( {n_A } \right)}$ is the expected number of neighbours of an interfering ED,
$\Xi$ is the proportion of EDs in the annulus within the contention of transmiter, expressed as $\Xi  = \int_0^{2{l_j}} {\left[ {1 - {F_{\rm K}}\left( {\frac{{{\mathcal{P} _0}}}{{{\mathcal{P} _{\rm tx}}g\left( x \right)}}} \right)} \right]} {f_{\mathfrak{R}}}\left( x \right)dx$,
where ${{\mathcal{P} _0}}$ is the detection threshold, $\mathcal{P}_{\rm tx}$ is the power consumption of an ED when transmitting the data and ${f_{\mathfrak{R}}}\left( x \right)$ is the distance distritution of two independent random EDs uniformly distributed inside a circle of radius ${\mathfrak{R}}$ \cite{1999An},
i.e., ${f_{\mathfrak{R}}}\left( x \right) = \frac{{4x}}{{\pi {{\mathfrak{R}}^2}}}\left[ {{{\cos }^{ - 1}}\left( {\frac{x}{{2{\mathfrak{R}}}}} \right) - \frac{x}{{2{\mathfrak{R}}}}\sqrt {1 - \frac{{{x^2}}}{{4{{\mathfrak{R}}^2}}}} } \right], 0 \le x \le 2\mathfrak{R}$.

\section{Performance Analyses} \label{sect:Performance_Analysis}
In this section, we comprehensively evaluate the performance of LoRa based WBANs in terms of BEP, success probability, coverage probability, energy efficiency, throughput, and system delay.
\subsection{BEP Analysis} \label{sect:Closed-form}

The received signal-to-noise ratio (SNR) is given by

\begin{align}
\label{eq:receive_snr}
\gamma  = \frac{{{E_s}/{T_s}}}{{B{N_0}}} {\left| h \right|^2} = \frac{{{E_s}}g(d)}{{{N_0} {2^{\nu}}}}  {\left| {{\hbar}} \right|^2}  = \frac{{\mathcal{P}_{\rm tx}}g(d)}{{N}}{\left| {{\hbar}} \right|^2},
\end{align}
where ${N} =  - 174 + \partial  + 10{\log _{10}}B$ dBm, $\partial $ is the noise figure of receiving equipment and is fixed for a particular hardware implementation as $6$ dB, and $\overline{\gamma}=\frac{{{E_s}}g(d)}{{{N_0} {2^{\nu}}}}$ is the average SNR.

The PDFs of power gains for the Rayleigh fading channel and  shadowing are respectively written as
\begin{equation}
\begin{aligned}
\label{eq:raypdf}
{p_\mathrm{ray}}\left( x  \right) = {e^{ - x }},
\end{aligned}
\end{equation}
\begin{equation}
\begin{aligned}
\label{eq:logpdf}
{p_\mathrm{log }}\left( x  \right) = \frac{1}{{\sqrt {2\pi } {\sigma _H}x }}{e^{ - \frac{{{{(\ln x  - {\mu _H})}^2}}}{{2\sigma _H^2}}}}.
\end{aligned}
\end{equation}

Using a gamma distribution to approximate the log-normal distribution \cite{778479}, one can obtain
\begin{equation}
\begin{aligned}
\label{eq:logpdfgamma}
{p_\mathrm{log }}\left( x \right) \approx \frac{1}{{\Gamma \left( \psi  \right)}}{\left( {\frac{\psi }{\varepsilon }} \right)^\psi }{x^{\psi  - 1}}{e^{ - x\frac{\psi }{\varepsilon }}},
\end{aligned}
\end{equation}
where $\Gamma \left ( \cdot   \right ) $ is the gamma function, $\psi  = \frac{1}{{{e^{\sigma _{_H}^2}} - 1}}$ and $\varepsilon  = {e^{{\mu _H}}}\sqrt {\frac{{\psi  + 1}}{\psi }} $.

Let $\beta = {\left| {{\hbar}} \right|^2} = {\left| {{h_\mathrm{ray}}} \right|^2}H $, according to \cite{9000517} and \cite{5403551}, the PDF of $\beta$ can be computed as
\begin{align}
\label{eq:rlogpdfaccurate}
{p_\beta }\left( z \right) \approx {p_\mathrm{ray}}\left( z \right) \times {p_\mathrm{log }}\left( z \right) = \frac{{{z^{\xi  - 1}}}}{{\Gamma \left( \xi  \right){\delta ^\xi }}}{e^{ - \frac{z}{\delta }}},
\end{align}
where $\xi  = \frac{1}{{2{e^{\sigma _H^2}} - 1}}$ and $\delta  = \left( {2{e^{\sigma _H^2}} - 1} \right){e^{{\mu _H}}}\sqrt {{e^{\sigma _H^2}}} $.

Using $\int_0^z {{x^t}{e^{ - \varsigma x}}} dx = {\varsigma ^{ - t - 1}}\mathcal{G} \left( {t + 1,\varsigma z} \right)$, the approximated cumulative distribution function (CDF) of $\beta$ can be calculated as
\begin{equation}
\begin{aligned}
\label{eq:rlogcdf}
{F_\beta }\left( z \right) = \int_{ - \infty }^z {{p_\beta }(x)dx \approx \frac{1}{{\Gamma \left( \xi  \right)}}} \mathcal{G} \left( {\xi ,\frac{z}{\delta }} \right),
\end{aligned}
\end{equation}
where $\mathcal{G} \left ( \cdot ,\cdot  \right ) $ is the lower incomplete gamma function.

According to the analysis in Sect.~\ref{sect:Signal Model for  LoRa Physical Layer}, the BEP of the LoRa system over a Rayleigh-lognormal fading channel with \textit{co-SF} interference can be given by
\begin{align}
\label{eq:ber}
{P_b} &= \frac{{{2^{\nu  - 1}}}}{{{2^\nu } - 1}} {P_e} \nonumber\\
&\approx \frac{1}{2} \left( {\frac{1}{{{2^\nu }}}  \left. {{P_e}} \right|\iota  = 0 + {\frac{{{2^\nu } - 1}}{{{2^\nu }}}} \left. {{P_e}} \right|\iota  \ne 0} \right)\nonumber\\
&\overset{\mathrm{\left ( a \right ) }}  \approx \frac{1}{2}  \left( {\frac{{{2^\nu } - 1}}{{{2^\nu }}} \left. {{P_e}} \right|\iota  \ne 0} \right) \nonumber\\
&\overset{\mathrm{\left ( b \right ) }}\approx \frac{1}{2}  \left( {1 - \left( {1 - P_e^{\left( \mathrm N \right)}} \right) \left( {1 - P_e^{\left( \mathrm I \right)}} \right)} \right) \nonumber\\
&\approx ~\frac{1}{2}  \left( {P_e^{\left( \mathrm N \right)} + \left( {1 - P_e^{\left( \mathrm N \right)}} \right)  P_e^{\left( \mathrm I \right)}} \right),
\end{align}
where $P_e$ represents the corresponding symbol error probability (SEP), $P_e^{\left( \mathrm N \right)}$ and $P_e^{\left( \mathrm I \right)}$ denote the SEP over a Rayleigh-lognormal fading channel in the case of no interference and \textit{co-SF} interference, respectively.
Here, in step $\mathrm{\left ( a \right ) }$, the BEP is assumed to be determined by the case of $\iota \ne 0$, i.e., the magnitude of the correlator output can be approximated as (\ref{eq:magnitude_output2}).
In step $\mathrm{\left ( b \right ) }$, ${\frac{{{2^\nu } - 1}}{{{2^\nu }}}}$ is approximated to be $1$, and ${\left. {{P_e}} \right|\iota  \ne 0}$ is expressed by $P_e^{\left( \mathrm N \right)}$ and $P_e^{\left( \mathrm I \right)}$.

Firstly, according to \cite{8392707}, $P_e^{\left( \mathrm N \right)}$ can be given by
\begin{align}
\label{eq:rlogpe}
&P_e^{\left( \mathrm N \right)} = \int\limits_0^\infty  {Q\left( {\sqrt {{2^{\nu + 1}}\overline{\gamma} x}   - \sqrt {2{H_{{2^{\nu}} - 1}}} } \right){p_{\beta} }\left( x  \right)} dx \nonumber\\
&  \approx \int\limits_0^\infty  {Q\left( {\sqrt {{2^{\nu  + 1}}\overline{\gamma} x}  - \sqrt {2{H_{{2^\nu } - 1}}} } \right)  } \frac{{{x^{\xi  - 1}}}}{{\Gamma \left( \xi  \right){\delta ^\xi }}}{e^{ - \frac{x}{\delta }}}dx,
\end{align}
where ${H_\varsigma  } \approx \ln \left( \varsigma  \right) + \frac{1}{{2\varsigma }} + 0.57722$.

Using the linear approximation and some mathematical calculations \cite{8861362}, $Q\left( {\sqrt {{2^{\nu  + 1}}\overline{\gamma} x}  - \sqrt {2{H_{{2^\nu } - 1}}} } \right)$ in (\ref{eq:rlogpe}) can be approximated as
\begin{align}
\label{eq:qfunclinear}
Q\!\left( \!\!{\sqrt {{2^{\nu  + 1}}\overline{\gamma} x} \! -\!\! \sqrt {2{H_{{2^\nu } - 1}}} } \right) \!\!\approx\!\! \left\{ {\begin{array}{*{20}{c}}
\!\!	1,&{x\! \le a\!\! +\!\! \frac{1}{{2b}}}\\
\!\!	{\frac{1}{2}\!\! +\! b\!\left( {x \!\!-\!\! a} \right)},&{a \!\!+\!\! \frac{1}{{2b}}\!\! < x <\!\! a \!\!-\!\! \frac{1}{{2b}}}\\
\!\!	0,&{x \ge a \!\!- \!\!\frac{1}{{2b}}}
	\end{array}} \right.,
\end{align}
where $a = \frac{{{H_{{2^\nu } - 1}}}}{{{2^\nu }\overline{\gamma} }}$ and $b =  - \frac{{{2^\nu }\overline{\gamma} }}{{2\sqrt {\pi {H_{{2^\nu } - 1}}} }}$.

Then, (\ref{eq:rlogpe}) can be written as
\begin{align}
\label{eq:rlogpe2}
P_e^{\left( \mathrm N \right)} \approx &\int\limits_0^{a + \frac{1}{{2b}}}  \frac{{{x^{\xi  - 1}}}}{{\Gamma \left( \xi  \right){\delta ^\xi }}}{e^{ - \frac{x}{\delta }}}dx + \nonumber\\
&\int\limits_{a + \frac{1}{{2b}}}^{a - \frac{1}{{2b}}} {\left( {\frac{1}{2} + b\left( {x - a} \right)} \right)}  \frac{{{x^{\xi  - 1}}}}{{\Gamma \left( \xi  \right){\delta ^\xi }}}{e^{ - \frac{x}{\delta }}}dx.
\end{align}

By merging the same parts of the integration interval, the closed-form approximation of $P_e^{\left( \mathrm N \right)} $ can be given in (\ref{eq:rlogpe3}).
\begin{figure*}[ht]
\begin{align}
\label{eq:rlogpe3}
P_e^{\left( \mathrm N \right)}& \approx \frac{{a\! +\! \frac{1}{{2b}}}}{{\Gamma \left( \xi  \right)}}\mathcal{G} \left( \!{\xi ,\frac{1}{\delta }\left( {a + \frac{1}{{2b}}} \right)} \!\right) + \frac{{a \!-\! \frac{1}{{2b}}}}{{\Gamma \left( \xi  \right)}}\mathcal{G} \left(\! {\xi ,\frac{1}{\delta }\left( {a - \frac{1}{{2b}}} \right)} \!\right) + \frac{{b\delta }}{{\Gamma \left( \xi  \right)}}\left[ {\mathcal{G} \left( \!{\xi  + 1,\frac{1}{\delta }\left( {a - \frac{1}{{2b}}}\! \right)} \right) - \mathcal{G} \left(\! {\xi  + 1,\frac{1}{\delta }\left( {a + \frac{1}{{2b}}} \right)} \!\right)} \right]\nonumber\\
&\approx \frac{{{H_{{2^\nu } - 1}} - \sqrt {\pi {H_{{2^\nu } - 1}}} }}{{{2^\nu }\overline{\gamma} \Gamma \left( \xi  \right)}}\mathcal{G} \left( {\xi ,\frac{1}{\delta }\left( {\frac{{{H_{{2^\nu } - 1}} - \sqrt {\pi {H_{{2^\nu } - 1}}} }}{{{2^\nu }\overline{\gamma} }}} \right)} \right) + \frac{{{H_{{2^\nu } - 1}} + \sqrt {\pi {H_{{2^\nu } - 1}}} }}{{{2^\nu }\overline{\gamma} \Gamma \left( \xi  \right)}}\mathcal{G} \left( {\xi ,\frac{1}{\delta }\left( {\frac{{{H_{{2^\nu } - 1}} + \sqrt {\pi {H_{{2^\nu } - 1}}} }}{{{2^\nu }\overline{\gamma} }}} \right)} \right) \nonumber\\
& - \frac{{{2^\nu }\overline{\gamma} \delta }}{{2\sqrt {\pi {H_{{2^\nu } - 1}}} \Gamma \left( \xi  \right)}}\left[ {\mathcal{G} \left( {\xi  + 1,\frac{1}{\delta }\left( {\frac{{{H_{{2^\nu } - 1}} + \sqrt {\pi {H_{{2^\nu } - 1}}} }}{{{2^\nu }\overline{\gamma} }}} \right)} \right) - \mathcal{G} \left( {\xi  + 1,\frac{1}{\delta }\left( {\frac{{{H_{{2^\nu } - 1}} - \sqrt {\pi {H_{{2^\nu } - 1}}} }}{{{2^\nu }\overline{\gamma} }}} \right)} \right)} \right].
\end{align}
	{\noindent} \rule[-10pt]{18cm}{0.05em}
\end{figure*}

\begin{figure*}[ht]
	\begin{align}
	\label{eq:rlogpeI}
	P_e^{\left( \mathrm I \right)} \!\!\approx \!\!\frac{{\sum\nolimits_{\tau  = 0}^{{2^{\nu  - 1}}} {\left[ {\frac{1}{{{2^\nu }}}  \sum\nolimits_{{I_1} = 0}^{{2^\nu } - 1} {\int_0^\infty  {\int_0^\infty  {Q\left( {\sqrt {{2^\nu }\overline{\gamma} {\beta _1}}  - \sqrt {{2^\nu }\overline{\gamma} {\beta _{\hat{k}}}} {\mathcal{U} _0}} \right)  {p_\beta }\left( {{\beta _1}} \right)  {p_\beta }\left( {{\beta _{\hat{k}}}} \right)d{\beta _1}d{\beta _{\hat{k}}}} } } } \right]} }}{{{2^{\nu  - 1}} + 1}}\!\! \approx \!\!\frac{{\sum\nolimits_{\tau  = 0}^{{2^{\nu  - 1}}} {\left[ {\frac{1}{{{2^\nu }}} \sum\nolimits_{{I_1} = 0}^{{2^\nu } - 1} {P_e^I\left| {_{{\mathcal{U} _0}}} \right.} } \right]} }}{{{2^{\nu  - 1}} + 1}}.
	\end{align}
	{\noindent} \rule[-10pt]{18cm}{0.05em}
\end{figure*}

Secondly, according to the deduced formula of the SEP over AWGN channel in \cite{8903531}, $P_e^{\left( \mathrm I \right)}$ can be calculated as (\ref{eq:rlogpeI}).
The double integral part in (\ref{eq:rlogpeI}) is denoted by $P_e^{\left( \mathrm I \right)} \left| {_{{\mathcal{U} _0}}} \right.$.

Since double integral is very complicated, we firstly integrate it with respect to $\beta_1$.
Let $A = \sqrt {{2^\nu }\overline{\gamma} {\beta _{\hat{k}}}} {\mathcal{U} _0}$, and according to the Hermite integration in \cite[Table 25.10]{1970Handbook}, one can obtain
\begin{align}
\label{eq:pe_integral1}
&\int\limits_0^\infty  {Q\left( {\sqrt {{2^\nu }\overline{\gamma} {\beta _1}}  - A} \right)  {p_\beta }\left( {{\beta _1}} \right)} d{\beta _1} \nonumber\\
&\approx \int\limits_0^\infty  {Q\left( {\sqrt {{2^\nu }\overline{\gamma} {\beta _1}}  - A} \right) \frac{{{\beta _1}^{\xi  - 1}}}{{\Gamma \left( \xi  \right){\delta ^\xi }}}{e^{ - \frac{{{\beta _1}}}{\delta }}}} d{\beta _1} \nonumber\\
&\approx \int\limits_{ - \infty }^{ + \infty } {Q\left( {\sqrt {{2^\nu }\overline{\gamma} {e^{{y_1}}}}  - A} \right)}  \frac{{{e^{{y_1}}}^{\left( {\xi  - 1} \right)}}}{{\Gamma \left( \xi  \right){\delta ^\xi }}}{e^{ - \frac{{{e^{{y_1}}}}}{\delta }}} {e^{{y_1}}}d{y_1} \nonumber\\
&\approx \sum\limits_{{w_1} = 1}^{{W_1}}\! {{\zeta _{w_1}}{e^{{y_1}^2}}} {e^{{y_1}\xi  - \frac{{{e^{{y_1}}}}}{\delta }}} \frac{1}{{\Gamma \left( \xi  \right)\!{\delta ^\xi }}}  Q\left( {\sqrt {{2^\nu }\overline{\gamma} {e^{{y_1}}}} \! - \!A} \right)\! +\! {O_{{W_1}}},
\end{align}
where $W_1$ is the number of sample points for approximation, ${y_1} = \ln {\beta _1}$ denotes the integral point, ${\zeta _{w_1}}$ denotes the weight factors and $O_{W_1}$ is the remainder which is approximate to $0$ as $W_1$ approaches infinity.

Substituting $A$ into (\ref{eq:pe_integral1}), and integrating (\ref{eq:pe_integral1}) with respect to $\beta_{\hat{k}}$, $P_e^{\left( \mathrm I \right)} \left| {_{{\mathcal{U} _0}}} \right.$ can be further calculated as
\begin{align}
\label{eq:pe_integralI1}
&P_e^{\left( \mathrm I \right)} \left| {_{{\mathcal{U} _0}}} \right. = \int\limits_0^\infty  \left[ {\sum\limits_{{w_1} = 1}^{{W_1}} {{\zeta _{w_1}}{e^{{y_1}^2}}}   {e^{{y_1}\xi  - \frac{{{e^{{y_1}}}}}{\delta }}}  \frac{1}{{\Gamma \left( \xi  \right){\delta ^\xi }}} } \right.\nonumber\\
&\left.{ Q\left( \!{\sqrt {{2^\nu }\overline{\gamma} {e^{{y_1}}}} \! - \!\sqrt {{2^\nu }\overline{\gamma} {\beta _{{\hat{k}}}}} {\mathcal{U} _0}} \!\right) \! +\! {O_{{W_1}}}} \right]   \frac{{{\beta _{{\hat{k}}}}^{\xi  - 1}}}{{\Gamma \left( \xi  \right){\delta ^\xi }}}{e^{ - \frac{{{\beta _{{\hat{k}}}}}}{\delta }}}d{\beta _{{\hat{k}}}}.
\end{align}

Let ${y_2} = \ln {\beta _{{\hat{k}}}}$, and similar to the processing of (\ref{eq:pe_integral1}), the closed-form approximation of $P_e^{\left( \mathrm I \right)} \left| {_{{\mathcal{U} _0}}} \right.$ can be given in (\ref{eq:pe_integral2}), where $W_2$ is the number of sample points for approximation, ${\zeta _{w_2}}$ denotes the weight factors and $O_{W_2}$ is the remainder which is approximate to $0$ as $W_2$ approaches infinity.
To make the approximation more accurate, ${W_1} = {W_2} = 20$ are adopted.

\begin{figure*}[ht]
	\begin{align}
	\label{eq:pe_integral2}
P_e^{\left( \mathrm I \right)}\!\left| {_{{\mathcal{U} _0}}} \right.\! = \!\frac{1}{{{{\Big( {\Gamma \left( \xi  \right){\delta ^\xi }} \Big)}^2}}}\sum\limits_{{w_2} = 1}^{{W_2}} {\sum\limits_{{w_1} = 1}^{{W_1}} {\Bigg[ {{\zeta _{{w_2}}}{e^{{y_2}^2}}{e^{{y_2}\xi  - \frac{{{e^{{y_2}}}}}{\delta }}}  \Big( {{\zeta _{{w_1}}}{e^{{y_1}^2}}{e^{{y_1}\xi  - \frac{{{e^{{y_1}}}}}{\delta }}}  Q\left( {\sqrt {{2^\nu }\Gamma {e^{{y_1}}}}  - \sqrt {{2^\nu }\Gamma {e^{{y_2}}}} {\mathcal{U} _0}} \right)\! + \!{O_{{W_1}}}} \Big)} \Bigg]} }  + {O_{{W_2}}}.
	\end{align}
	{\noindent} \rule[-10pt]{18cm}{0.05em}
\end{figure*}

Finally, by combining (\ref{eq:ber}), (\ref{eq:rlogpe3}), (\ref{eq:rlogpeI}) and (\ref{eq:pe_integral2}), we can obtain the closed-form expression of $P_b$.

\begin{table}[]
	\begin{center}
		\caption{SF Specific Threshold ${q_{\nu}}$ ($B = 125$ KHz).}
		\begin{tabular}{ c c c c }
			\hline
			Annulus  &  $\nu$ & \makecell[c]{SNR $q_{\nu}$ \\(dBm)} & \makecell[c]{Range \\(m)}   \\
			\hline
			$1$ &  $7$  &  $-6$  &  $l_0$--$l_1$\\
			$2$ &  $8$  &  $-9$  &  $l_1$--$l_2$\\
			$3$ &  $9$  &  $-12$  &  $l_2$--$l_3$\\
			$4$ &  $10$  &  $-15$  &  $l_3$--$l_4$\\
			$5$ &  $11$  &  $-17.5$  &  $l_4$--$l_5$\\
			$6$ &  $12$  &  $-20$  &  $l_5$--$R$\\
			
			\hline
		\end{tabular}
		\label{table:sf}	
	\end{center}
	
\end{table}

\subsection{Coverage Probability} \label{sect:Coverage Probability}


In interference-free scenarios, the system is affected by fading and the noise. If the received SNR is below the reception threshold ${q_{\nu}}$, which is shown in Table~\ref{table:sf}, that allows successful detection, the node can not  connect to the gateway.
By definiton,  the connection probability is formulated as

\begin{equation}
\begin{aligned}
\label{eq:connectionprobability}
{\rm P_{SNR}} = \mathbb{P} \left[ {\gamma  \ge  {q_{\nu}}|d_1} \right], {d_1} \in \left[ {{l_{j - 1}},{l_j}} \right].
\end{aligned}
\end{equation}

To find the transmission range when using LoRa signals for WBAN, we need to  find the relationship between the distance $d_1$ and the connection probability.
By using $\gamma$ in (\ref{eq:receive_snr}), the connection probability  can be computed as
\begin{align}
\label{eq:pcon1}
&{\rm P_{SNR}}\left( {{d_1}} \right) = \mathbb{P}\left[ {\frac{{{\mathcal{P}_{\rm tx}}}}{N}  {{\left| {\hbar_1} \right|}^2}  g\left( {{d_1}} \right) \ge {q_{\nu}}} \right]\nonumber\\
&= \mathbb{P}\left[ {{{\left| {\hbar_1} \right|}^2} \ge \frac{{N  {q_{\nu}}}}{{{\mathcal{P}_{\rm tx}}  g\left( {{d_1}} \right)}}} \right] \nonumber\\
&= 1 - {F_{\beta}}\left( {\frac{{N  {q_{\nu}}}}{{{\mathcal{P}_{\rm tx}}  g\left( {{d_1}} \right)}}} \right)\nonumber\\
&\approx 1 - \frac{1}{{\Gamma \left( \xi  \right)}}\mathcal{G} \left( {\xi ,\frac{{N{q_\nu }}}{{{\mathcal{P}_{\rm tx}}g\left( {{d_1}} \right)\delta }}} \right) .
\end{align}
where $\mathbb{P} \left [ \cdot \right ] $ is the notation for calculating probability.

We can see from (\ref{eq:pcon1}) that the connection probability is affected by the distance $d_1$ and the transmitted power ${\mathcal{P}_{\rm tx}}$.
Increasing the transmitted power at the same distance can achieve higher connection probability and thus improving the communication reliability.


Considering an interference scenario, since we use the assumption that different SFs are perfectly orthogonal to one another, the system is able to provide full protection for concurrent transmissions from different SFs.
Moreover, except for the case where the smallest SF is used by a very large number of EDs, we only need to consider the dominant \textit{co-SF} interferer.
And the transmitted power of EDs with the same SF signals are supposed to be equal.
Due to the capture effect of the LoRa, the stronger signals will suppress weaker signals received at the same time \cite{9018210}.
We define the strongest interferer $\hat{k}$ as

\begin{align}
\label{eq:strongestk}
{\hat{k} } &= \arg \mathop{\max}_{k>1} \left\{ {{\mathcal{P}_{\rm tx}}\chi _k^{\nu}{{\left| {{h_k}} \right|}^2}} \right\} \nonumber
\\&= \arg \mathop{\max}_{k>1} \left\{ {{\mathcal{P}_{\rm tx}}\chi _k^{\nu}{{\left| {\hbar_k} \right|}^2}  g\left( {{d_k}} \right)} \right\},
\end{align}
where $d_k$ and ${\hbar_k}$ denote the distance and the channel gain between the $k$-th interfering node and the gateway, respectively.
Hence, the received SIR of the desired ED under dominant \textit{co-SF} interference is given by
\begin{align}
\label{eq:defineSIR}
\rho  = \frac{{{\mathcal{P}_{\rm tx}}{{\left| {\hbar_1} \right|}^2}g\left( {{d_1}} \right)}}{{{\hat{\mathcal{I} } }}},
\end{align}
where ${\hat{\mathcal{I} } } = {\mathcal{P}_{\rm tx}}{\left| {\hbar_{{\hat{k} }}} \right|^2}g\left( {{d_{{\hat{k} }}}} \right)$ is the dominant \textit{co-SF} interference.

After we find ${\hat{k} }$, the SIR success probability can be written as
\begin{align}
\label{eq:pcon2}
&{\rm P_{SIR}} \left( {{d_1} } \right)  = \mathbb{P} \left[ {\left. {\frac{{{{\left| {\hbar_1} \right|}^2}g\left( {{d_1}} \right)}}{{{{\left| {\hbar_{\hat{k} }} \right|}^2}g\left( {{d_{{\hat{k} }}}} \right)}} \ge \theta } \right|{d_1}} \right] \nonumber \\&= {\mathbb{E} _{{{\left| {\hbar_1} \right|}^2}}}\left[\! {\mathbb{P} \left[\! {\left. {{X_{{\hat{k} }}} \le \frac{{{{\left| {\hbar_1} \right|}^2}g\left( {{d_1}} \right)}}{\theta }} \right|{{\left| {\hbar_1} \right|}^2},{d_1}} \right]} \right],
\end{align}
where $\mathbb{E} \left [ \cdot \right ]$ represents the statistical expectation, ${X_{{\hat{k} }}} = {\left| {\hbar_{{\hat{k} }}} \right|^2}g\left( {{d_{{\hat{k} }}}} \right)$, and $\theta = 1$ dB is the SIR threshold \cite{8430542}.
If the desired signal is $\theta$ dB stronger than any other signal received simultaneously, no collision occurs.

To find the SIR success probability under the strongest interferer, let ${X_i} = {\left| {\hbar_i} \right|^2}g\left( {{d_i}} \right)$. According to the previous analysis, since ${d_1} \in \left[ {{l_{j - 1}},{l_j}} \right]$, we can obtain that $\left| {{\mathcal{V} _{d_1}}} \right| = \pi \left( {l_j^2 - l_{j - 1}^2} \right)$ and $\mathcal{V}_{d_1}\subset\mathcal{V}$. Moreover, the PDF of $d_i$, which is defined as the distance between the gateway and the randomly selected ED within the same annulus $\mathcal{V}_{d_1}$, can be written as ${f_{{d_i}}}\left( x \right) = \frac{{2\pi x}}{{\left| {{\mathcal{V}_{d_1}}} \right|}}$. Hence, the PDF of $g\left( {{d_i}} \right)$ can be calculated as

\begin{align}
\label{eq:pdfd}
{f_{g\left( {{d_i}} \right)}}\left( x \right) &= \left| {\frac{d}{{dx}}{g^{ - 1}}\left( x \right)} \right|{f_{{d_i}}}\left( {{g^{ - 1}}\left( x \right)} \right) \nonumber \\&= \frac{{2\pi  \times {{10}^{ - \frac{{2{P_L}\left( {{d_0}} \right)}}{{10n}}}}{x^{ - \frac{2}{n} - 1}}}}{{n\left| {{\mathcal{V}_{d_1}}} \right|}},
\end{align}
where $g\left( {{l_j}} \right) \le x \le g\left( {{l_{j - 1}}} \right)$. Using (\ref{eq:rlogpdfaccurate}), the approximated closed-form PDF of $X_i$ can be computed as

\begin{align}
\label{eq:pdfxi}
&{f_{{X_i}}}\left( z \right) = {f_{g\left( {{d_i}} \right)}}\left( z \right) \otimes {p_{\beta}}\left( z \right) \nonumber \\&= \int\limits_{g\left( {{l_j}} \right)}^{g\left( {{l_{j - 1}}} \right)} {\frac{1}{x}} {f_{g\left( {{d_i}} \right)}}\left( x \right){p_{\beta}}\left( {\frac{z}{x}} \right)dx \nonumber \\& \approx \int\limits_{g\left( {{l_j}} \right)}^{g\left( {{l_{j - 1}}} \right)} {\frac{1}{x}} \frac{{2\pi  \times {{10}^{ - \frac{{2{P_L}\left( {{d_0}} \right)}}{{10n}}}}}}{{n\left| {{{\cal V}_{{d_1}}}} \right|}}{x^{ - \frac{2}{n} - 1}}\frac{1}{{\Gamma \left( \xi  \right){\delta ^\xi }}}{\left( {\frac{z}{x}} \right)^{\xi  - 1}}{e^{ - \frac{z}{{x\delta }}}}dx \nonumber \\
& \approx \frac{{2\pi \!\! \times\!\! {{10}^{ - \frac{{2{P_L}\left( {{d_0}} \right)}}{{10n}}}}}}{{n\left| {{{\cal V}_{{d_1}}}} \right|\Gamma\! \left( \xi  \right)}}{\left( {\frac{1}{\delta }} \right)^{ - \frac{2}{n}}}{z^{ - \frac{2}{n} - 1}}\left. {\mathcal{G} \left( {\frac{2}{n} + \xi ,\frac{z}{{\delta g\left( x \right)}}} \right)} \right|_{x = {l_{j - 1}}}^{x = {l_j}},
\end{align}
where the symbol $ \otimes $ stands for convolution and $z>0$. By integrating (\ref{eq:pdfxi}) and exchanging the order of integration, the CDF of $X_i$ can be given in (\ref{eq:cdfxi}).

\begin{figure*}[ht]
	\begin{align}
	\label{eq:cdfxi}
{F_{{X_i}}}\left( z \right) = \int\limits_0^z {{f_{{X_i}}}\left( x \right)dx}  = \frac{{\pi   {{10}^{ - \frac{{2{P_L}\left( {{d_0}} \right)}}{{10n}}}}}}{{\left| {{{\cal V}_{{d_1}}}} \right|\Gamma \left( \xi  \right)}}  \left. {\left[ {g{{\left( x \right)}^{ - \frac{2}{n}}}\mathcal{G} \left( {\xi ,\frac{z}{{\delta g\left( x \right)}}} \right) - {{\left( {\frac{z}{\delta }} \right)}^{ - \frac{2}{n}}}\mathcal{G} \left( {\frac{2}{n} + \xi ,\frac{z}{{\delta g\left( x \right)}}} \right)} \right]} \right|_{x = {l_{j - 1}}}^{x = {l_j}}.
	\end{align}
	{\noindent} \rule[-10pt]{18cm}{0.05em}
\end{figure*}

According to the order statistics, the CDF of the maximun interference is ${F_{{X_{{\hat{k} }}}}}\left( z \right) = {\mathbb{E} _{{\mathcal{M}}} }\left[ {{{\left[ {{F_{{X_i}}}\left( z \right)} \right]}^{{\mathcal{M}}} }} \right]$, where the sample size ${\mathcal{M}}$ is a Poisson distrubuted random variable with mean $\upsilon  = {\lambda _u}\left| {{\mathcal{V} _{d_1}}} \right|$ and $\upsilon$ is the expected number of concurrently transmitting EDs in the same SF annulus ${\mathcal{V} _{d_1}}$.
Since the value of ${\mathcal{M}}$ is an integer from 0 to infinity, according to the total probability theorem, one has
\begin{align}
\label{eq:F_xk}
{F_{{X_{{\hat{k} }}}}}\left( x \right) = \sum\limits_{k = 0}^\infty  {\frac{{{\upsilon ^k}{e^{ - \upsilon }}}}{{k!}}} {\left[ {{F_{{X_i}}}\left( x \right)} \right]^k}.
\end{align}

Using the series formula ${e^x} = \sum\limits_{k = 0}^\infty  {\frac{{{x^k}}}{{k!}}} $ and deconditioning on the channel power gain ${\left| {\hbar_1 } \right|^2}$, one can obtain
\begin{align}
\label{eq:pcon2finally}
{\rm P_{SIR}} \left( {{d_1}} \right) = {e^{ - \upsilon }}\int\limits_0^\infty  {{e^{\upsilon {F_{{X_i}}}\left( {\frac{{zg\left( {{d_1}} \right)}}{\theta }} \right)}}} {p_{\beta}}\left( z \right)dz.
\end{align}

We can see from (\ref{eq:pcon2finally}) that SIR success probability is not only affected by the distance $d_1$, but also the mean of Poisson distrubuted random variable $\upsilon$, which is related to the channel access mechanisms.

The success probability of the proposed system can be computed as
\begin{align}
\label{eq:pjointsuc}
{\rm{P_{{\mathop{\rm joint}} }}} \left( {{d_1}} \right) \approx {\rm P_{SNR}} \left( {{d_1}} \right)  {\rm P_{SIR}} \left( {{d_1}} \right).
\end{align}

Hence, through averaging over $\mathcal{V}$, the coverage probability of the specific ED is given by
\begin{align}
\label{eq:pcoverage}
{\rm P_c} = \frac{2}{{{R^2}}}\sum\limits_j {\int_{{l_{j - 1}}}^{{l_j}} {{\rm{P_{{\mathop{\rm joint}} }}}\left( {{x}} \right)} } {x} d{x}.
\end{align}

\subsection{Energy Efficiency} \label{sect:Energy Efficiency}
Energy efficiency is defined as the ratio between  the number of successfully decoded bits and consumed energy of the system per unit, which represents
the efficiency of the system's use of energy resources.
In the sleep mode it can be assumed that no ED is working, hence the energy consumption can be negligible. Thus, the energy efficiency for an ED at a distance $x$ from the gateway can be written as
\begin{align}
\label{eq:energy_efficiency}
{\eta _{\rm EE}}\left( x \right) = \frac{{{\rm{P_{{\mathop{\rm joint}} }}} \left( x \right)  {N_{bp}}}}{{{E_u}}},
\end{align}
where ${E_u}$ is the energy required for transmitting a message to the gateway, which is related to the channel access mechanisms in the network.

According to \cite{9018210}, for the P-ALOHA protocol, one has
\begin{align}
\label{eq:EEPALOHA}
{E_{_\mathrm{P\raisebox{0mm}{-}ALOHA}}} = {\mathcal{P} _{\rm tx}}{T_o }.
\end{align}

For the S-ALOHA protocol, let $\mathcal{P}_{\rm rx}$ represent the power consumption of an ED when receiving the data, one has
\begin{align}
\label{eq:EESALOHA}
{E_{_\mathrm{S\raisebox{0mm}{-}ALOHA}}}  = {\mathcal{P} _{\rm tx}}{T_o } + {\mathcal{P} _{\rm rx}}  {T_B}\left( {\frac{{T_o }}{{\alpha {T_{\rm SYN}}}}} \right),
\end{align}
where the synchronization is maintained by the gateway sending periodic beacons of $T_B$ duration every interval $T_{\rm SYN}$.

For the NP-CSMA, one has
\begin{align}
\label{eq:EECSMA}
{E_{_\mathrm{NP\raisebox{0mm}{-}CSMA}}}  = {\mathcal{P} _{\rm tx}}{T_o } + {\mathcal{P} _{\rm rx}} {T_{\rm CAD}}\left[ {\frac{{\overline {E\left( n_A \right)}}}{{1 - {e^{ - \overline {E\left( n_A \right)}}}}}} \right],
\end{align}
where ${T_{\rm CAD}}$ is the  channel activity detection (CAD) duration, and $\overline {E\left( n_A \right)} = \lambda p  \pi \left( {l_j^2 - l_{j - 1}^2} \right)\Xi $ is the expected number of active neighbors.

Through averaging over $\mathcal{V}$, the average energy efficiency of the proposed system can be computed as
\begin{align}
\label{eq:energy_efficiency_cov}
\overline{\eta}_{EE} = \frac{2}{{{R^2}}}\sum\limits_j \int_{{l_{j - 1}}}^{{l_j}}
{\eta _{\rm EE}}\left( x \right) {x} d{x}.
\end{align}

\subsection{Throughput} \label{sect:Throughput}
According to  Section~\ref{sect:system_model}, one has a Poisson distributed EDs with the offered traffic $G$, where $G = \alpha \overline{\mathcal{N}}$. The throughput of the proposed system for the $j$-th annulus for all MAC protocols is given by
\begin{align}
\label{eq:throughput}
{\mathcal{CT}_j}  =  \frac{2G}{{{R^2}}}\int_{{l_{j - 1}}}^{{l_j}} {{\rm{P_{{\mathop{\rm joint}} }}} \left( {{x}} \right)} {x} d{x}.
\end{align}

Hence, the average throughput for the area $\mathcal{V}$ is calculated as
\begin{align}
\label{eq:throughput_cov}
\overline{\mathcal{CT} } = \sum\limits_j {\mathcal{CT}_j} .
\end{align}

\subsection{Delay} \label{sect:Delay}
The delay is defined as the number of transmissions required to successfully transmit a packet. In this paper, we do not consider the processing delay and propagation delay. The delay of the proposed system for the $j$-th annulus for all MAC protocols is given by
\begin{align}
\label{eq:delay}
{D_j} = \frac{1}{{\frac{2}{{{R^2}}}\int_{{l_{j - 1}}}^{{l_j}} {{\rm{P_{{\mathop{\rm joint}} }}}\left( x \right)  xdx} }}.
\end{align}

Hence, the average delay for the area $\mathcal{V}$ is calculated as
\begin{align}
\label{eq:delay_cov}
\overline D = \sum\limits_j {{D_j}} .
\end{align}

\begin{table}
	\begin{center}
		\caption{Some Parameters for LoRa.}
		\begin{tabular}{ c| c| c }
			\hline
			Parameters  &  Symbol & Value \\
			\hline
			Bandwidth &  $B$  &  125 KHz\\
			\hline
			Carrier Frequency &  ${f_c}$  &  868 MHz  \\			
			\hline
			Transmit Power &  ${\mathcal{P} _{\rm tx}}$  &  14 dBm  \\
			\hline				
			Power Consumption for Received Data &  ${\mathcal{P} _{\rm rx}}$  &  15.18 mW  \\
			\hline						
			Pathloss Exponent &  $n$  &  2.8  \\
			\hline
			Pathloss & ${P_L}\left( {{d_0}} \right)$& 49.6 dB	\\		
			\hline
			Message Preamble & $N_{sp}$   &  8 symbols \\
			\hline
			Activity Factor & $\alpha$& 0.33\%	\\
			\hline
			Coding Rate   & $CR$ & 4/8 \\
			\hline
			Payload & ${N_{bp}}$ & 10 bytes\\
			\hline
			Guard Interval &  ${T_g}$ & 10.24 ms \\
			\hline
			Detection Threshold  & ${{\mathcal{P} _0}}$ & -150 dB \\						
			\hline		
			\makecell[c]{Beacon Preamble +\\Payload Synchronization Interval}  &   ${T_{\rm SYN}}$ & 128~s   \\
			\hline
			CAD Duration  & ${T_{\rm CAD}}$  & 2 symbols\\
			\hline
			
		\end{tabular}
		\label{table:loraparameters}	
	\end{center}
\end{table}

\section{Results and Discussion} \label{sect:Results_and_Discussion}
In this section, the analytical results are validated by using Monte-Carlo simulations.
In this paper, some typical parameters are shown in Table~\ref{table:loraparameters}.

\begin{figure}[t]
	\center
	\includegraphics[width=3.8in,height=2.52in]{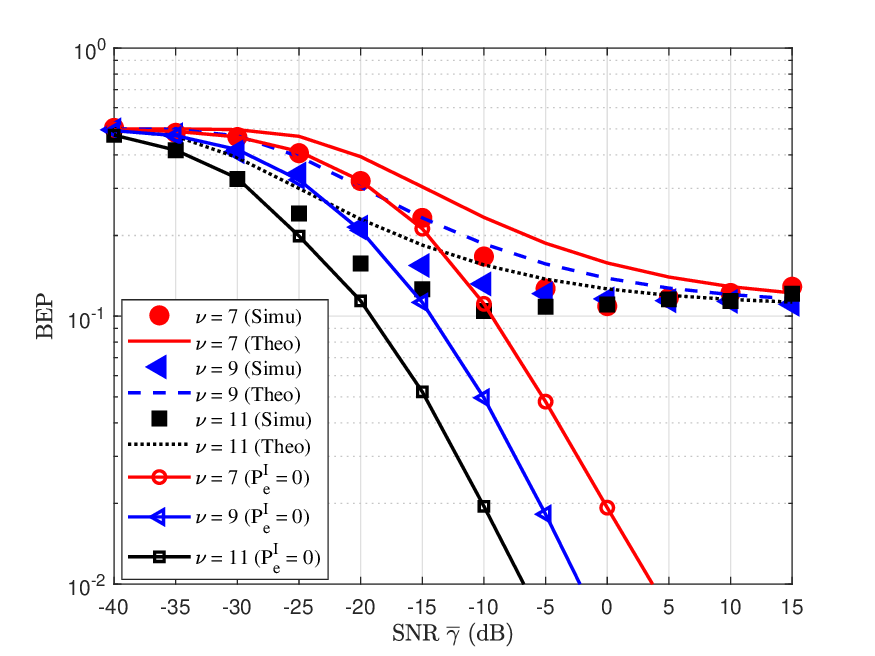}
	\caption{The theoretical and simulated BEP curves of the proposed system over Rayleigh-lognormal fading channel with no interference and \textit{co-SF} interference for $\nu = 7, 9, 11$, where ${\sigma_{dB}} = 8$ dB and $\rho = 6$ dB.}
	\label{fig:fig3}
	\vspace{-0.45cm}
\end{figure}

\subsection{BEP} \label{sect:ber}
Fig.~\ref{fig:fig3} presents the theoretical and simulated BEP curves over Rayleigh-lognormal fading channel with no interference and \textit{co-SF} interference for $\nu$ = $7$, $9$, and $11$, respectively, where the standard deviation ${\sigma_{dB}}$ is $8$ dB and the SIR $\rho$ is $6$ dB.
From this figure, under \textit{co-SF} interference, there is a small gap between the theoretical closed-form results and the simulated results due to the use of approximations in the derivation.
The curves with $P_e^{\left( \mathrm I \right)} = 0$ represent the case with no interference.
Moreover, it can be observed from this figure that considering the \textit{co-SF} interference, the BEP performance of the LoRa system deteriorates seriously.
With increasing the value of $\nu$, better performance can be obtained when $\Gamma < 5$ dB, while the BEP can be regarded as a fixed value when $\Gamma \ge 5$ dB.
In addition, at the same BEP, the SNR gaps between $\nu = 7$ and $\nu = 9$, $\nu = 9$ and $\nu = 11$ are about $5$ dB.

\begin{figure}[t]
	\center
	\includegraphics[width=3.8in,height=2.52in]{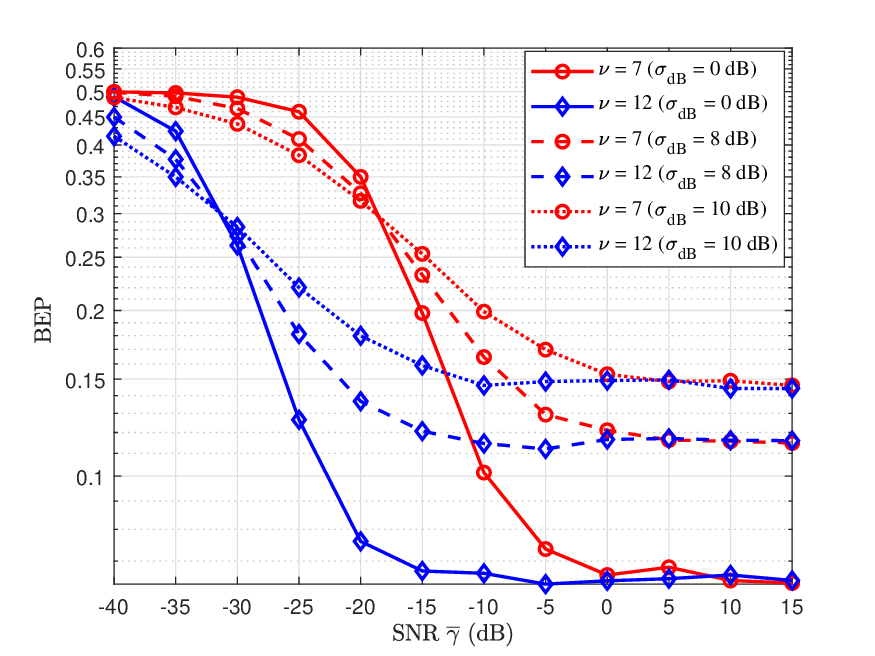}
	\caption{The simulated BEP curves of the proposed system with various values of ${\sigma_{dB}}$ and $\nu$ over Rayleigh-lognormal fading channel with \textit{co-SF} interference, where $\rho = 6$ dB.}
	\label{fig:fig4}
	\vspace{-0.45cm}
\end{figure}

Fig.~\ref{fig:fig4} shows the simulated BEP curves of the proposed system with various values of ${\sigma_{dB}}$ and $\nu$ over Rayleigh-lognormal fading channel with \textit{co-SF} interference.
From this figure, at the same value of $\nu$, better performance can be achieved when decreasing the standard deviation.
For example, at a BEP of $0.2$ and $\nu = 7$, the SNR gap is about $2.5$~dB between $\sigma_{dB} = 10$ dB and $\sigma_{dB} = 8$ dB, and between $\sigma_{dB} = 8$ dB and $\sigma_{dB} = 0$ dB.
Secondly, it can be seen that compared to the unshadowed Rayleigh fading channel, the BEP the proposed system over shadowed Rayleigh fading channel is obviously affected by the shadowing.
When $\Gamma \ge 5$ dB, the BEP gap between unshadowed ($\sigma_{dB} = 0$ dB) and shadowed ($\sigma_{dB} = 8$ dB) Rayleigh fading channel is about $0.12$.
Furthermore, regardless of the value of $\sigma_{dB}$, if the value of $\sigma_{dB}$ remains the same, the gain for increasing the value of  $\nu$ from $7$ to $12$ are almost constant and is about $12.5$ dB when $\Gamma < 5$ dB.

\begin{figure}[t]
	\center
	\includegraphics[width=3.8in,height=2.52in]{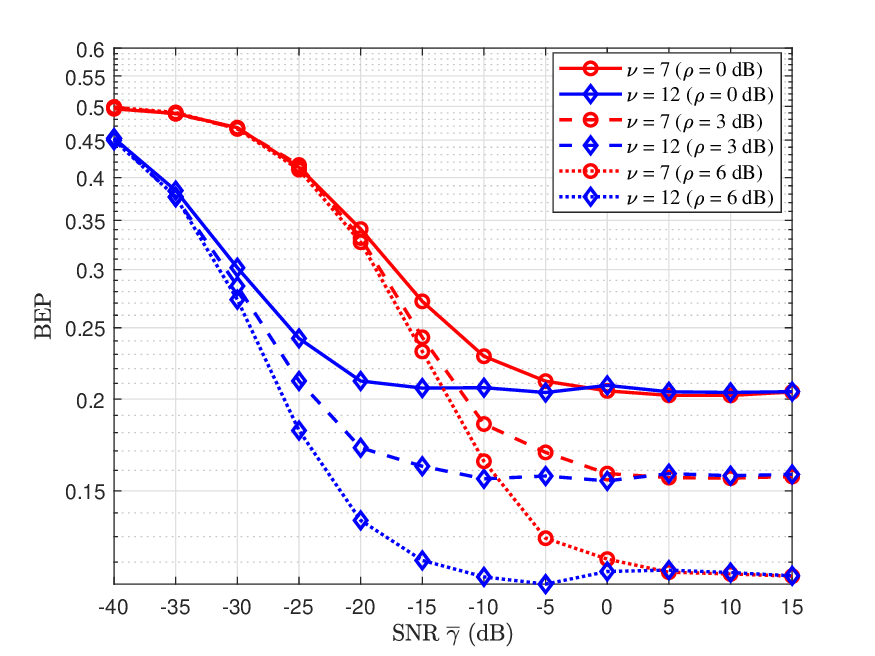}
	\caption{The simulated BEP curves of the proposed system with various values of ${\rho}$ and $\nu$ over Rayleigh-lognormal fading channel with \textit{co-SF} interference, where $\sigma_{dB} = 8$ dB.}
	\label{fig:fig5}
	\vspace{-0.45cm}
\end{figure}

Fig.~\ref{fig:fig5} shows the simulated BEP curves of the proposed system with various values of ${\rho}$ and $\nu$ over Rayleigh-lognormal fading channel with \textit{co-SF} interference.
From this figure, at the same value of $\nu$, the BEP can be effectively improved by increasing the SIR.
For example, when $\Gamma \ge 5$ dB, the BEP gap is $0.04$ between $\rho = 0$ dB and $\rho = 3$ dB, and it is $0.15$ between $\rho = 3$ dB and $\rho = 6$ dB.

Based on the above discussions, it can be found that BEP performance of the LoRa system  can be improved through increasing the values of $\nu$ and $\rho$, and reducing the value of $\sigma_{dB}$.
To reduce the BEP in practical applications, we can increase the SF and the SIR in the environments with low shadow effect, e.g., hilly/moderate-to-heavy tree density ($\sigma_{dB} = 2.3$~dB), hilly light tree density or flat/moderate-to-heavy tree density ($\sigma_{dB} = 3.0$~dB) and flat/light tree density ($\sigma_{dB} = 1.6$~dB)\cite{5071219}.
However, in the mountain ($\sigma_{dB} = 11.9$~dB) and canyon ($\sigma_{dB} = 10.13$~dB)\cite{9169667}, etc., reducing the shadow effect by adjusting the position of the antenna and increasing the SIR by controlling the power are also effective methods to obtain significant performance gain.

\begin{figure}[t]
	\begin{center}
		\subfigure[]
		{
			\centering
			\includegraphics[width=3.8in,height=2.52in]{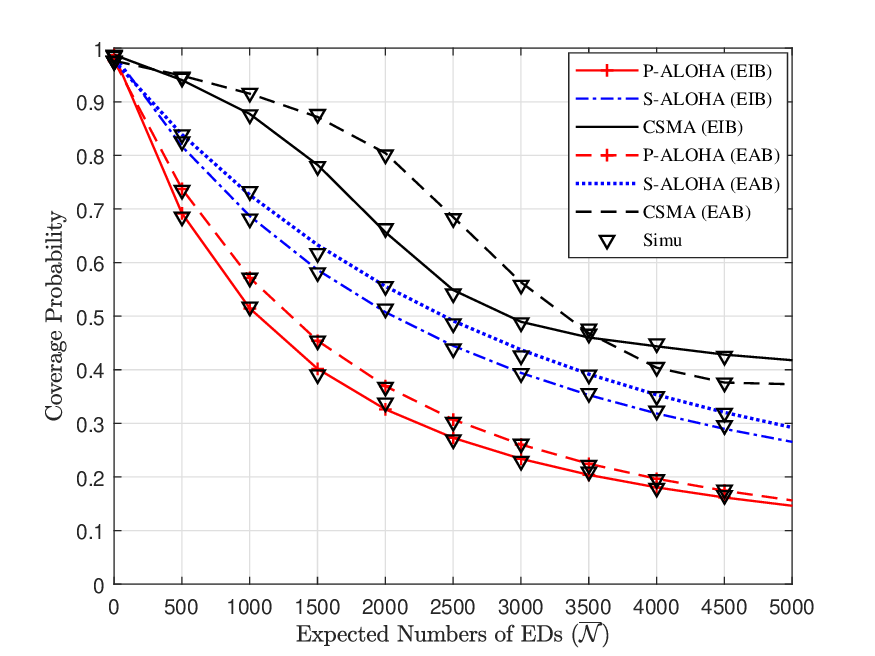}
		}
		\subfigure[]
		{
			\centering
			\includegraphics[width=3.8in,height=2.52in]{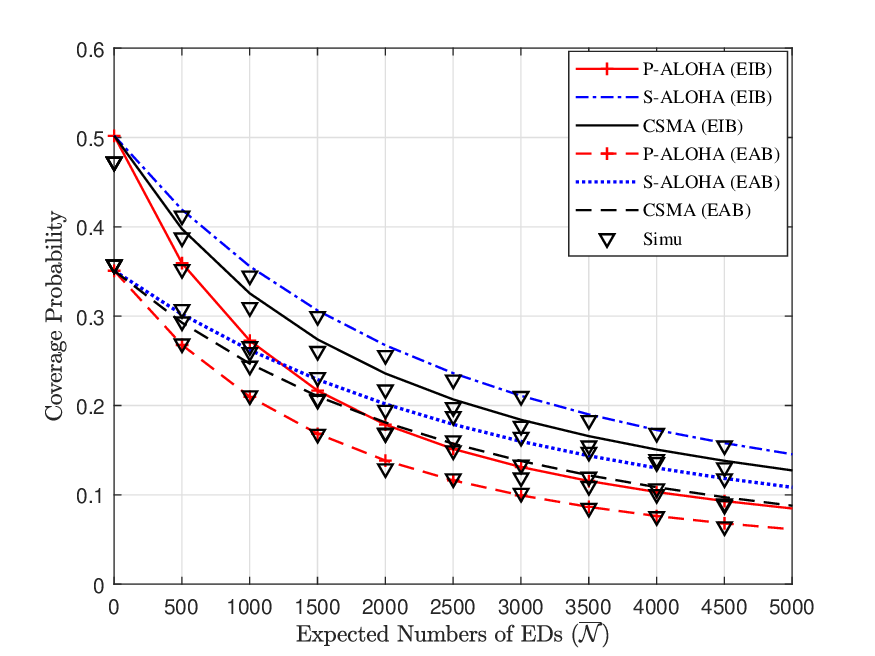}
		}
		\caption{Coverage probability with respect to average number of EDs for different channel access protocols. (a) $R = 1$ km. (b) $R = 6$ km.  \label{fig:fig6}}
	\end{center}
\end{figure}

\begin{figure}[t]
	\begin{center}
		\subfigure[]
		{
			\centering
			\includegraphics[width=3.8in,height=2.52in]{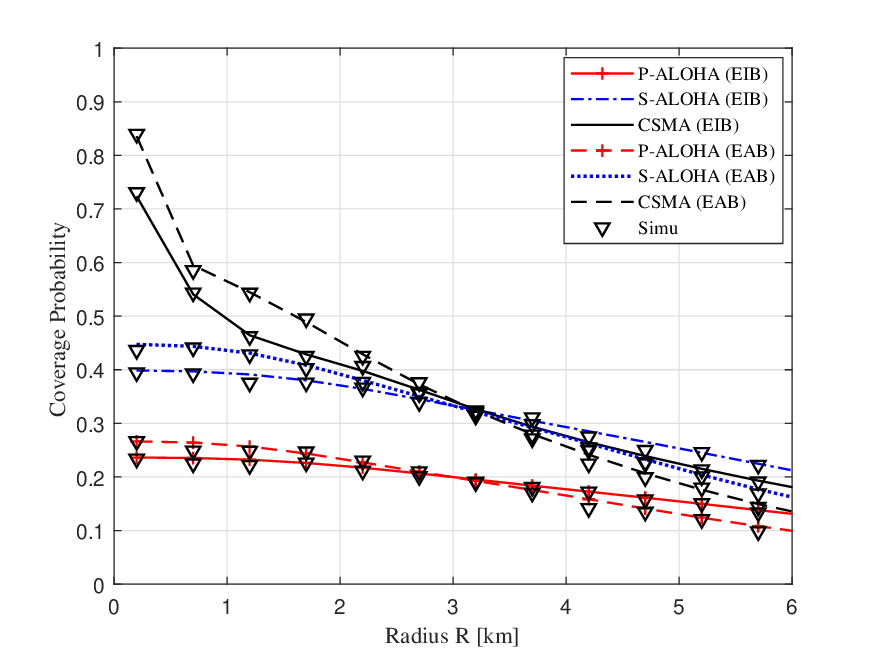}
		}
		\subfigure[]
		{
			\centering
			\includegraphics[width=3.8in,height=2.52in]{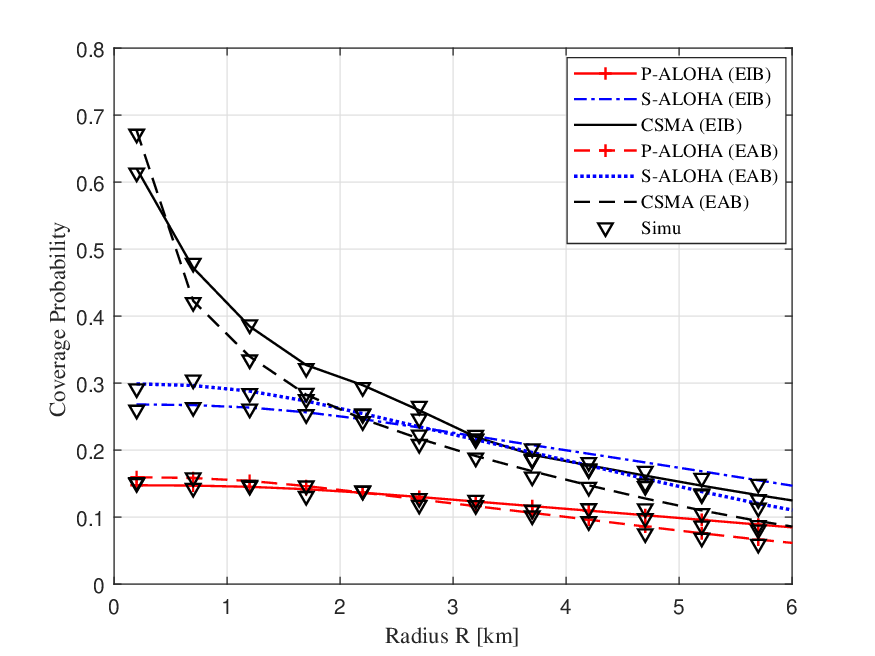}
		}
		\caption{Coverage probability with repsect to network radius $R$ for different channel access protocols. (a) $\overline {\mathcal{N}} = 3000$. (b) $\overline {\mathcal{N}} = 5000$.  \label{fig:fig7}}
	\end{center}
\end{figure}

\begin{figure}[t]
	\begin{center}
		\subfigure[]
		{
			\centering
			\includegraphics[width=3.8in,height=2.52in]{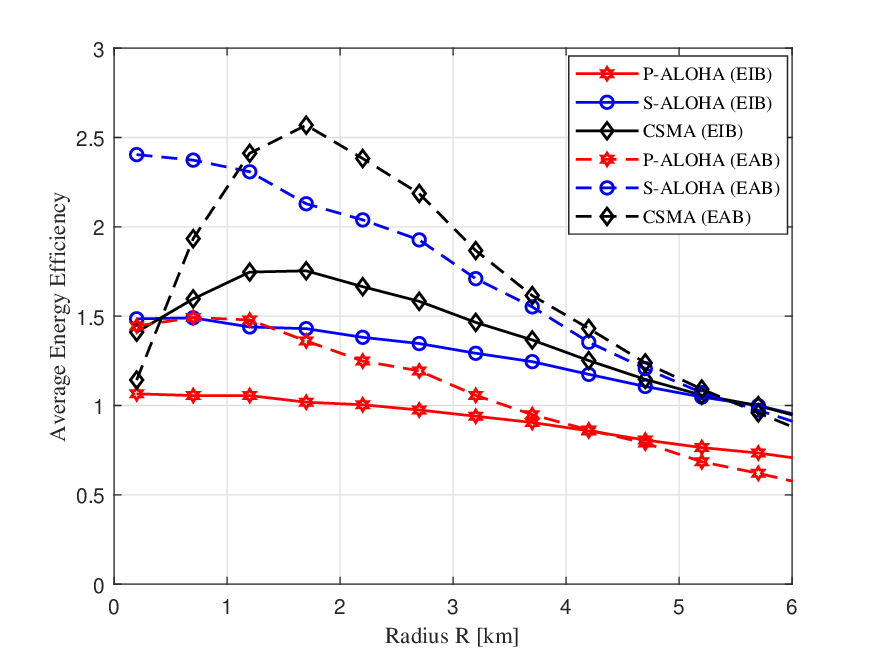}
		}
		\subfigure[]
		{
			\centering
			\includegraphics[width=3.8in,height=2.52in]{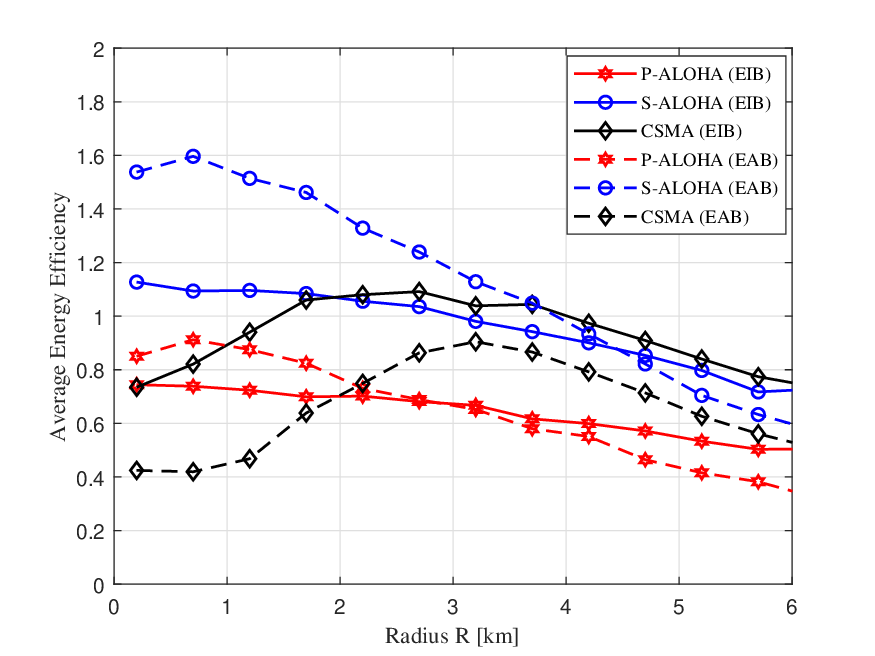}
		}
		\caption{Average energy efficiency with respect to network radius $R$  for different channel access protocols. (a) $\overline {\mathcal{N}} = 3000$. (b) $\overline {\mathcal{N}} = 5000$.  \label{fig:fig8}}
	\end{center}
\end{figure}

\begin{figure}[t]
	\begin{center}
		\subfigure[]
		{
			\centering
			\includegraphics[width=3.8in,height=2.52in]{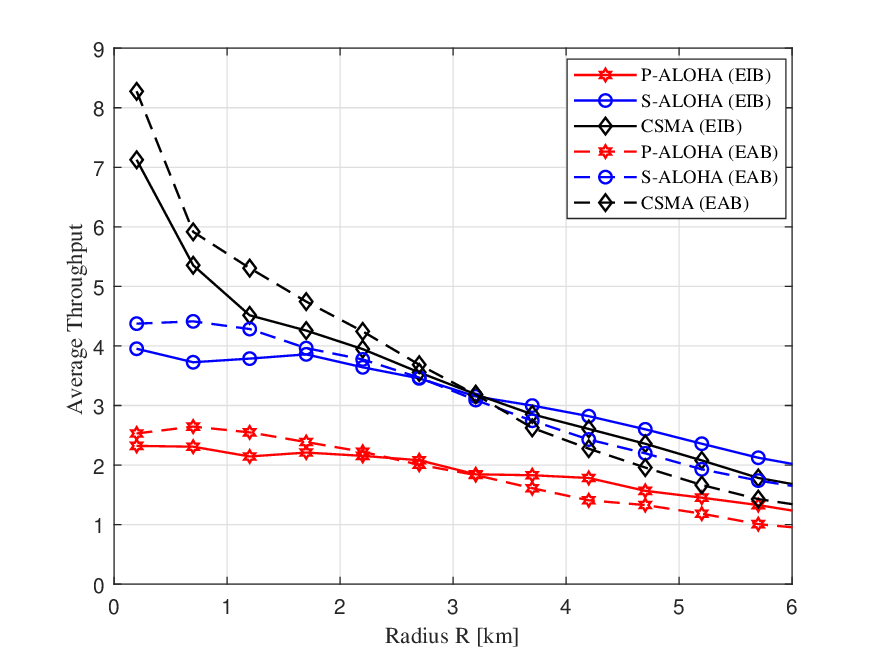}
		}
		\subfigure[]
		{
			\centering
			\includegraphics[width=3.8in,height=2.52in]{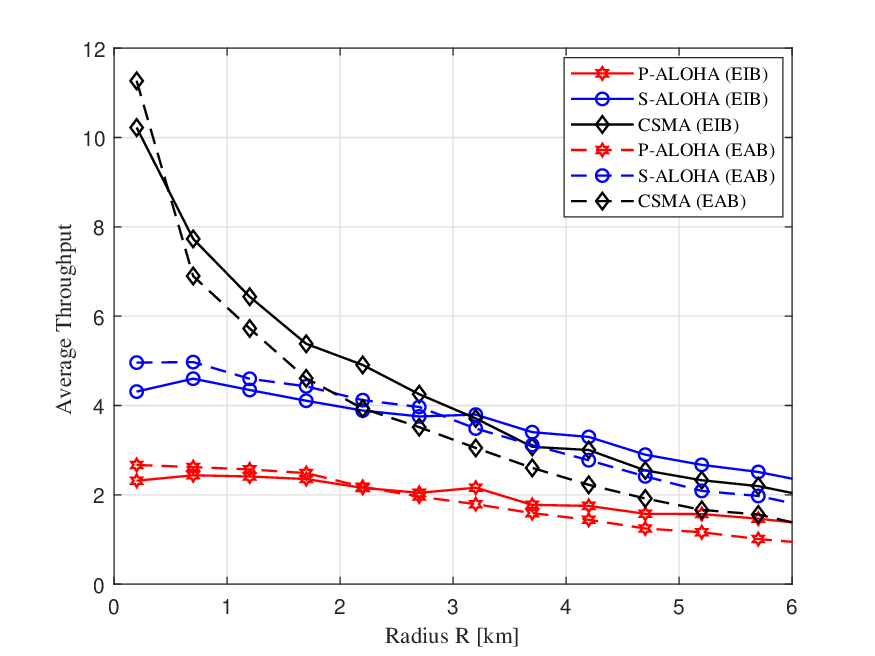}
		}
		\caption{Average throughput with respect to network radius $R$ for different channel access protocols. (a) $\overline {\mathcal{N}} = 3000$. (b) $\overline {\mathcal{N}} = 5000$.  \label{fig:fig9}}
	\end{center}
\end{figure}

\begin{figure}[t]
	\begin{center}
		\subfigure[]
		{
			\centering
			\includegraphics[width=3.8in,height=2.52in]{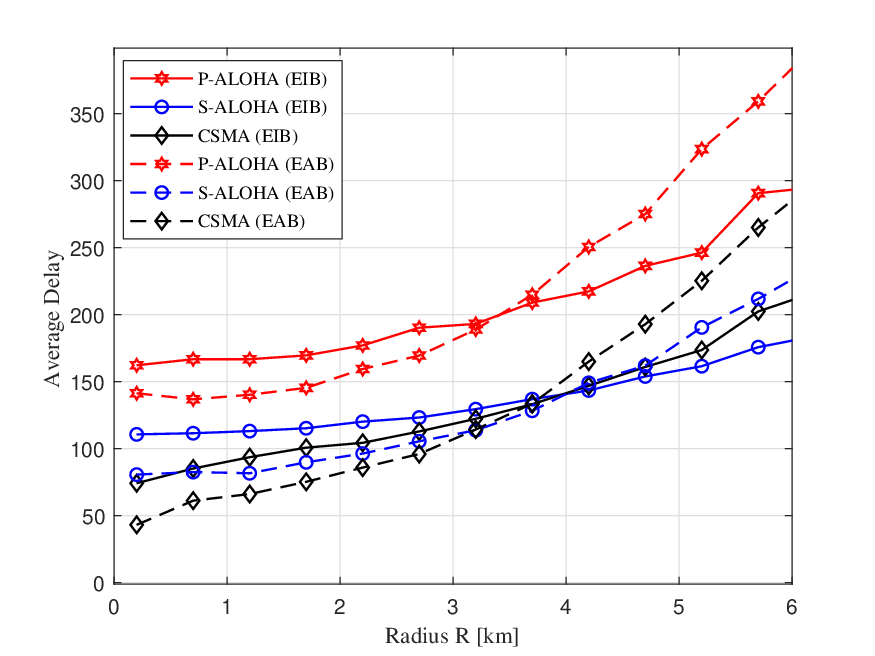}
		}
		\subfigure[]
		{
			\centering
			\includegraphics[width=3.8in,height=2.52in]{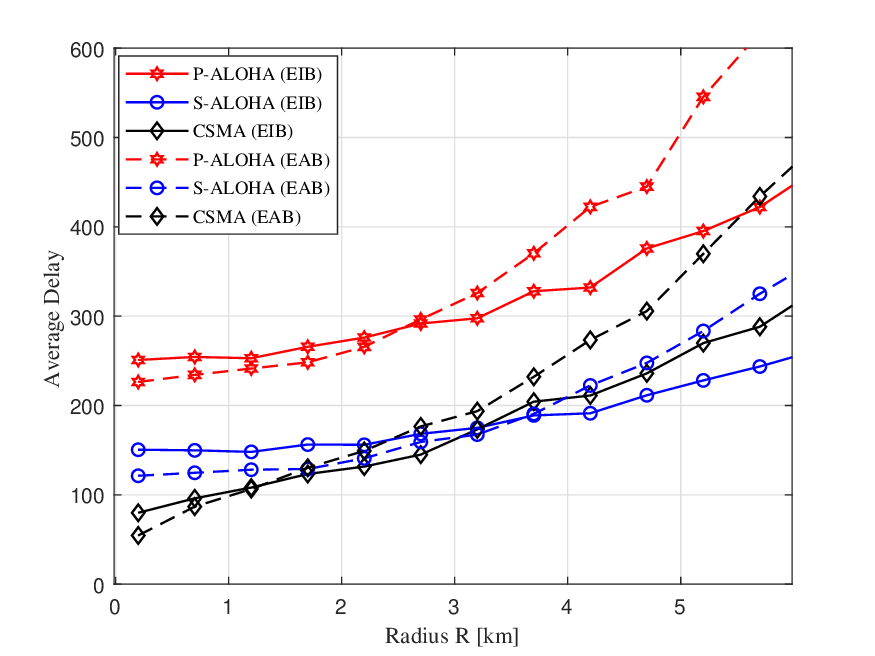}
		}
		\caption{Average delay with respect to network radius $R$ for different channel access protocols. (a) $\overline {\mathcal{N}} = 3000$. (b) $\overline {\mathcal{N}} = 5000$.  \label{fig:fig10}}
	\end{center}
\end{figure}

\subsection{Coverage Probability} \label{sect:pcov}

Fig.~\ref{fig:fig6} shows the coverage probability of the proposed system with respect to average number of EDs for the network radius $R = 1$ km and $6$ km.
Simulation results verify the theoretical derivations in (\ref{eq:pcon1}), (\ref{eq:pcon2finally}) and (\ref{eq:pjointsuc}).
It can be observed from Fig.~\ref{fig:fig6} that the coverage probability decreases with the increase of $\overline {\mathcal{N}}$ due to the increase of interfering sources.
As shown in Fig.~\ref{fig:fig6} (a), for a small value of $R$, the NP-CSMA protocol can access more number of EDs than the P-ALOHA and S-ALOHA protocol.
In addition, the number of EDs that can be accessed by using the EAB scheme is greater than that of the EIB scheme for the three protocols.
However, as shown in Fig.~\ref{fig:fig6} (b), for a large value of $R$,  more EDs can be accessed via the EIB scheme for each protocol and the S-ALOHA protocol can access more number of EDs.
Therefore, the network radius $R$ is an important factor for the choice of SF allocation schemes and the three MAC protocols.

Fig.~\ref{fig:fig7} shows the coverage probability with respect to $R$ for the three MAC protocols, where $\overline {\mathcal{N}} = 3000$, and $\overline {\mathcal{N}} = 5000$.
It can seen from these figures that the coverage probability decreases with the increase of the network radius $R$.
Moreover, it can be observed that for small value of the network radius $R$, the EAB scheme has a higher coverage probability than the EIB one and the NP-CSMA protocol has a higher coverage probability than the P-ALOHA and S-ALOHA protocols.
Furthermore, with the increase of the network radius $R$, the coverage probability of the EAB scheme decreases more than that of the EIB scheme, and the coverage probability of the NP-CSMA protocol decreases more than that of the S-ALOHA protocols.
After a specific value of the network radius $R$, the coverage probability of the EIB scheme exceeds the EAB scheme, the coverage probability of the S-ALOHA protocols exceeds the NP-CSMA protocol.
With increasing the number of EDs, the coverage probability of the EIB scheme exceeds that of the EAB scheme at a smaller value of the network radius $R$.
Consequently, to obtain higher coverage probability, for a small value of $R$, the EAB scheme performs better than the EIB scheme, and the NP-CSMA protocol is more efficient than the P-ALOHA and S-ALOHA protocols. For a large value of $R$, the EIB scheme and the S-ALOHA protocol could be adopted.


\subsection{Energy Efficiency} \label{Energy Efficiency}

Fig.~\ref{fig:fig8} shows the average energy efficiency curves of the proposed system with different values of $R$, where $\overline {\mathcal{N}} = 3000$ and $5000$.
It can be found that when the number of EDs is fixed, there exists the value of $R$ to reach the optimal average energy efficiency for the NP-CSMA protocol.
For example, when $\overline {\mathcal{N}} = 3000$, the NP-CSMA protocol has an optimal value of the average energy efficiency, i.e., $R \approx 1.8$ km shown in Fig.~\ref{fig:fig8} (a).
From Fig.~\ref{fig:fig8} (a) and (b), it can be seen that for the P-ALOHA and S-ALOHA protocol, the EAB scheme performs better than the EIB scheme for a small value of $R$ regardless of the value of $\overline {\mathcal{N}}$.
For the NP-CSMA protocol, the EAB scheme performs better than  the EIB scheme in a certain range of $R$ when $\overline {\mathcal{N}} = 3000$, while for $\overline {\mathcal{N}} = 5000$, better result can be achieved by using the EIB scheme regardless of the value of $R$.
Therefore, from the perspective of energy efficiency, in practical applications, the EAB scheme with the S-ALOHA protocol is suitable for a small value of $R$, while for a large value of $R$, with the increase of $\overline {\mathcal{N}}$, the EIB scheme with the NP-CSMA protocol can be utilized.

\subsection{Throughput} \label{Channel Throughput}

Fig.~\ref{fig:fig9} shows the average throughput
of the three MAC protocols for different values of $R$, where $\overline {\mathcal{N}} = 3000$ and $5000$.
It can be seen that to obtain higher throughput, the EAB scheme with the NP-CSMA protocol can be adopted for small network radius, while the EIB scheme with the S-ALOHA protocol can be used for large network radius.
The conclusion is the same as that of the coverage probability from Fig.~\ref{fig:fig7}.

\subsection{Delay} \label{Delay}

Fig.~\ref{fig:fig10} shows the average delay of the three MAC protocols for different values of $R$, where $\overline {\mathcal{N}} = 3000$ and $5000$.
From Fig.~\ref{fig:fig10}, it can be found that the increase of the network radius leads to an increase in the delay for a fixed number of EDs.
By comparing Fig.~\ref{fig:fig10} (a) with Fig.~\ref{fig:fig10} (b), increasing the number of EDs with a fixed network radius also leads to an increase in delay.
From the perspective of delay, the EIB scheme with the S-ALOHA can be used for large network radius, otherwise the EAB scheme with NP-CSMA protocol can be adopted.

\section{Conclusion} \label{sect:Conclusion}
In this paper, the performance of the LoRa system for WBAN has been investigated from PHY and MAC layers over Rayleigh-lognormal fading channel with \textit{co-SF} interference.
The closed-form BEP expression of the LoRa system under Rayleigh-lognormal fading channel and \textit{co-SF} interference has been derived.
The results show that increasing the value of SF $\nu$ and SIR $\rho$, and decreasing the value of standard deviation $\sigma_{dB}$ are effective ways to resist the shadowed effect in different environments.
Moreover, the performance of the P-ALOHA, S-ALOHA and NP-CSMA protocols for the LoRa based WBAN has been analyzed in terms of coverage probability, energy efficiency, throughput and system delay.
Furthermore, the performance of the EIB and EAB schemes for the LoRa based WBAN has also been investigated.
From the theoretical and simulated results, it can be found that to obtain higher coverage probability and throughput, and lower delay, the EAB scheme with the S-ALOHA protocol can be adopted for small network radius, while the EIB scheme with the NP-CSMA protocol is suggested for large network radius.
To achieve higher energy efficiency, the EAB scheme with the S-ALOHA protocol can be chosen for small network radius, while for large network radius, the EAB scheme with the NP-CSMA protocol and the EIB scheme with the NP-CSMA protocol are suitable for a small number of EDs and a large number of EDs, respectively.
Thanks to the these advantages, LoRa communication can be considered as a promising candidate for the WBAN.




\bibliographystyle{IEEEtran}


\end{document}